\documentclass[aps,prl,twocolumn,superscriptaddress]{revtex4-2}

\usepackage{graphicx}
\usepackage{amsmath,amssymb}
\usepackage{color}
\usepackage{bm,bbm,bbold}
\usepackage{multirow}

\usepackage[bookmarks=true,colorlinks=true,urlcolor=blue,linkcolor=blue,citecolor=blue,breaklinks]{hyperref}

\usepackage{booktabs}

\usepackage{color}
\hypersetup{
     colorlinks   = true,
     citecolor    = blue
}

\usepackage{ulem}

\begin{document}

\sloppy
\title{Flat-band optical phonons in twisted bilayer graphene}

\author{Emmanuele Cappelluti}
\affiliation{Istituto di Struttura della Materia, CNR (ISM-CNR), 34149 Trieste, Italy}
\author{Jose Angel Silva-Guill\'en}
\email{josilgui@gmail.com}
\affiliation{Instituto Madrile\~no de Estudios Avanzados, IMDEA Nanociencia, Calle Faraday 9, 28049, Madrid, Spain}
\author{Habib Rostami}
\affiliation{Department of Physics, University of Bath, Claverton Down, Bath BA2 7AY, United Kingdom}
\affiliation{Nordita, KTH Royal Institute of Technology and Stockholm University,
Hannes Alfv\'ens v\"ag 12, 10691 Stockholm, Sweden}
\author{Francisco Guinea}
\email{paco.guinea@gmail.com}
\affiliation{Instituto Madrile\~no de Estudios Avanzados, IMDEA Nanociencia, Calle Faraday 9, 28049, Madrid, Spain}
\affiliation{Donostia International Physics Center, Paseo Manuel de Lardiz\'abal 4, 20018, San Sebasti\'an, Spain; and Ikerbasque, Basque Foundation for Science, 48009, Bilbao, Spain}

\begin{abstract}

Twisting bilayer sheets of graphene have been proven to be an efficient way to manipulate the electronic Dirac-like properties, resulting in flat bands at magic angles. 
Inspired by the electronic model, we develop a continuum model for the lattice dynamics of twisted bilayer graphene and 
we show that a remarkable band flattening applies to almost all the high-frequency in-plane lattice vibration modes, including the valley Dirac phonon, valley optical phonon, and zone-center optical phonon bands. 
Utilizing an approximate approach, we estimate small but finite magic angles at which a vanishing phonon bandwidth is expected. 
In contrast to the electronic case, the existence of a restoring potential prohibits the emergence of a magic angle in a more accurate modeling. 
The predicted phonon band-flattening is highly tunable by the twist angle and this strong dependence is directly accessible by spectroscopic tools.

\end{abstract}

\maketitle
{\it Introduction.}
The exotic electronic, optical, and lattice properties of graphene have been enriched
in the past few years by the additional possibility of manipulating two graphene layers with a finite twist angle.
In twisted bilayer graphene (TBG), a complex phase diagram, including superconductivity, a Mott insulating phase, and a novel topology of the electronic bands,
has been revealed \cite{cao_nat18,cao2018}. A key ingredient in this scenario is the 
existence
of a non-trivial electronic structure with very narrow bandwidth, also known as flat-bands, at the so-called magic angle~ \cite{SCVPB10,DeTramblyLaissardiere2010a,BM11} has been analyzed using schemes based on either tight binding models \cite{SCVPB10,DeTramblyLaissardiere2010a}
or continuum models \cite{LPN07,M11}. 

Nevertheless, along with the investigation focused on the electronic properties, a large interest has also recently arisen
concerning the effects of twist on the lattice dynamics.
The phonon spectrum in TBG has been studied theoretically~\cite{CNB13}, and experimentally~\cite{JC13}. 
Optical \cite{WMM18} and acoustical \cite{LBWB19,WHS19} phonons have been investigated as possible origins of the observed superconductivity.
A particular high-energy optical mode at the K and K$^\prime$ points has been extensively studied in TBG \cite{ATF19,AF20,BF22},
as it gives rise to flat moir\'e bands, and it couples strongly to electrons.
These modes are also currently thought to be responsible for the
remarkable D and 2D features in Raman spectroscopy of single-layer
and multilayer graphene \cite{ferrari00,Petal04,Fetal06,basko07,B08}.
Concerning the possibility of a strong twist-driven renormalization
of the phonon dispersion, calculations based on models of elastic systems have also been carried out.
In Ref.~\cite{RPCS20} the emergence of a flat-band associated with out-of-plane flexural modes
was shown. Similar results for the out-of-plane lattice modes 
were predicted for twisted ``artificial''  graphene systems \cite{Detal20}.
In-plane lattice modes at the K and K$^\prime$ points,
also characterized by Dirac physics,
appear however as well, and are even more interesting.
On the one hand, these modes were initially associated with
the onset of the D and 2D Raman features \cite{matthews99,thomsen00}.
On the other hand, the same modes, in the presence of a symmetry breaking of the sublattices
as in h-BN or in transition-metal dichalcogenides (TMDs),
can host chiral content that enforces fundamental
selection rules \cite{zhang14,zhang15,zhu2018,chen2019,rgc22}.
In this scenario, it is worth mentioning that flat-bands
have been also predicted in moir\'e structures of twisted two-dimensional TMDs~\cite{SWZX21,MML22}.

\begin{figure*}[t]
\includegraphics[width=0.9\textwidth]{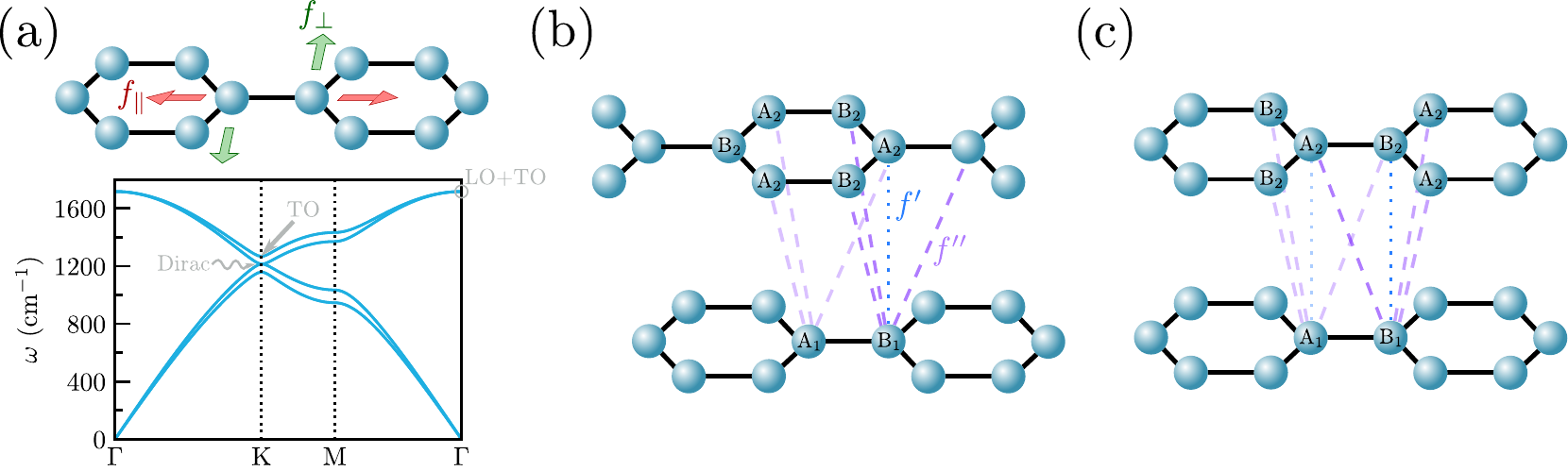}
\caption{Force-constant model for the untwisted cases.
(a) Single-layer graphene (top) and its phonon dispersion calculated using the model. Only the elastic coupling $f$ (solid black lines) between
nearest neighbor atoms $f$ is retained.
The colored arrows denote the lattice displacements coupled with the two
elastic components $f_\parallel$ (red) and $f_\perp$ (green).
(b),(c) AB and AA bilayer graphene, respectively. 
The two further interlayer elastic coupling are shown, $f^\prime$ (dotted blue lines) and $f^{\prime\prime}$ (dashed purple lines).}
\label{f-struct}
\end{figure*} 

In this Letter, we investigate the effect of twist on the 
main high-energy (optical) modes at the high-symmetry points
$\Gamma$ and K of the phonon spectrum of TBG, with a special focus on the Dirac-like in-plane lattice modes at K.
Using a force-constant (FC) model and a proper generalization of the continuum approach for the lattice phonon modes,
we show that: ($i$) in-plane Dirac-phonons
undergo upon twist a strong renormalization 
of the effective dispersion giving rise to flat-bands,
in a similar way as Dirac-like electrons do;
($ii$) a ``magic'' angle, where the dispersion of these modes approaches zero,
can be analytically predicted,
and numerically observed,
at twist angles remarkably larger than the ones
required for the existence of flat bands in the electronic spectrum.
Furthermore, we show that the appearance of flat bands
is also predicted (at smaller twist angles)
for the high-frequency transverse optical (TO) phonon at K 
and for the longitudinal-optical/transverse-optical (LO/TO) modes
at the $\Gamma$ point, rationalizing thus
the numerical results of Ref. \cite{ATF19}.

{\it The model.}
A suitable continuum model for the lattice dynamics
of TBG is derived from a FC model.
Following the well-known scheme,
we first construct the proper Hamiltonian for the single-layer,
and for the representative limit cases of AA and AB bilayer stacking.
The lattice dynamics for the twisted system is further obtained
by including the appropriate tunneling
between a ${\bf q}$ vector in one layer
with a ${\bf q}+{\bf Q}_\nu$ vector in the other layer,
where ${\bf Q}_\nu$ are the characteristic tunnelling momenta,
just as for the electronic case.
In order to focus on the physics of the Dirac phonons,
we restrict our model to in-plane lattice displacements responsible
for the Dirac modes, defining a 8-fold Hilbert basis,  $u_{\alpha,i}({\bf q})$,
corresponding to the lattice displacements of the 4 atoms
in the $x$-$y$ space.
Here $i=x,y$ are the Cartesian indices and $\alpha=$ A$_1$, B$_1$,
A$_2$, B$_2$ labels the atoms in the sublattice A, B
in layer 1, 2.

The phonon band structure is thus
obtained by the solution of the secular equation:
\begin{eqnarray}
M\, \hat{\omega}^2({\bf q}) 
\cdot
{\bf u}({\bf q})
&=&
{\bf \hat{K}}({\bf q})\cdot {\bf u}({\bf q})
,
\end{eqnarray}
where $M$ is the carbon mass,
$\hat{\omega}^2({\bf q})$ the diagonal matrix
of the square frequencies,
and ${\bf \hat{K}}({\bf q})$ the dynamical matrix
that takes into account the elastic couplings
between different carbon atoms.
In order to provide the clearest analytical insight
on the manipulation of the Dirac lattice modes,
we include the minimum set of FC parameters
preserving the relevant physics.
More explicitly, in single-layer we include elastic coupling
only between in-plane nearest neighbor atoms,
described by two parameters, $f_\parallel$, and $f_\perp$,
ruling the relative radial and in-plane tangential lattice displacements
between neighbor atoms at interatomic distance $a$ (see Fig. \ref{f-struct}a).
The coupling between different layers in the AA and AB structure
is thus described by two more kinds of
elastic forces (see Figs.~\ref{f-struct}b,c) : $f^\prime_\perp$ connecting vertically two 
atoms atop each other at the distance $c$; and $f^{\prime\prime}$, connecting
atoms in different layers at distance $R=\sqrt{a^2+c^2}$,
with the relevant components $f^{\prime\prime}_\parallel$
and $f^{\prime\prime}_\perp$, governing respectively the relative in-plane
longitudinal and transverse displacement of two atoms
with respect to their joining vector.
The resulting dynamical matrix can thus be written as:
\begin{eqnarray}
\hat{\bf K}({\bf q})
&=&
\hat{\bf K}^{f}({\bf q})
+
\hat{\bf K}^{f^\prime}({\bf q})
+
\hat{\bf K}^{f^{\prime\prime}}({\bf q})
.
\end{eqnarray}
\begin{figure*}[t]
    \centering
    \includegraphics{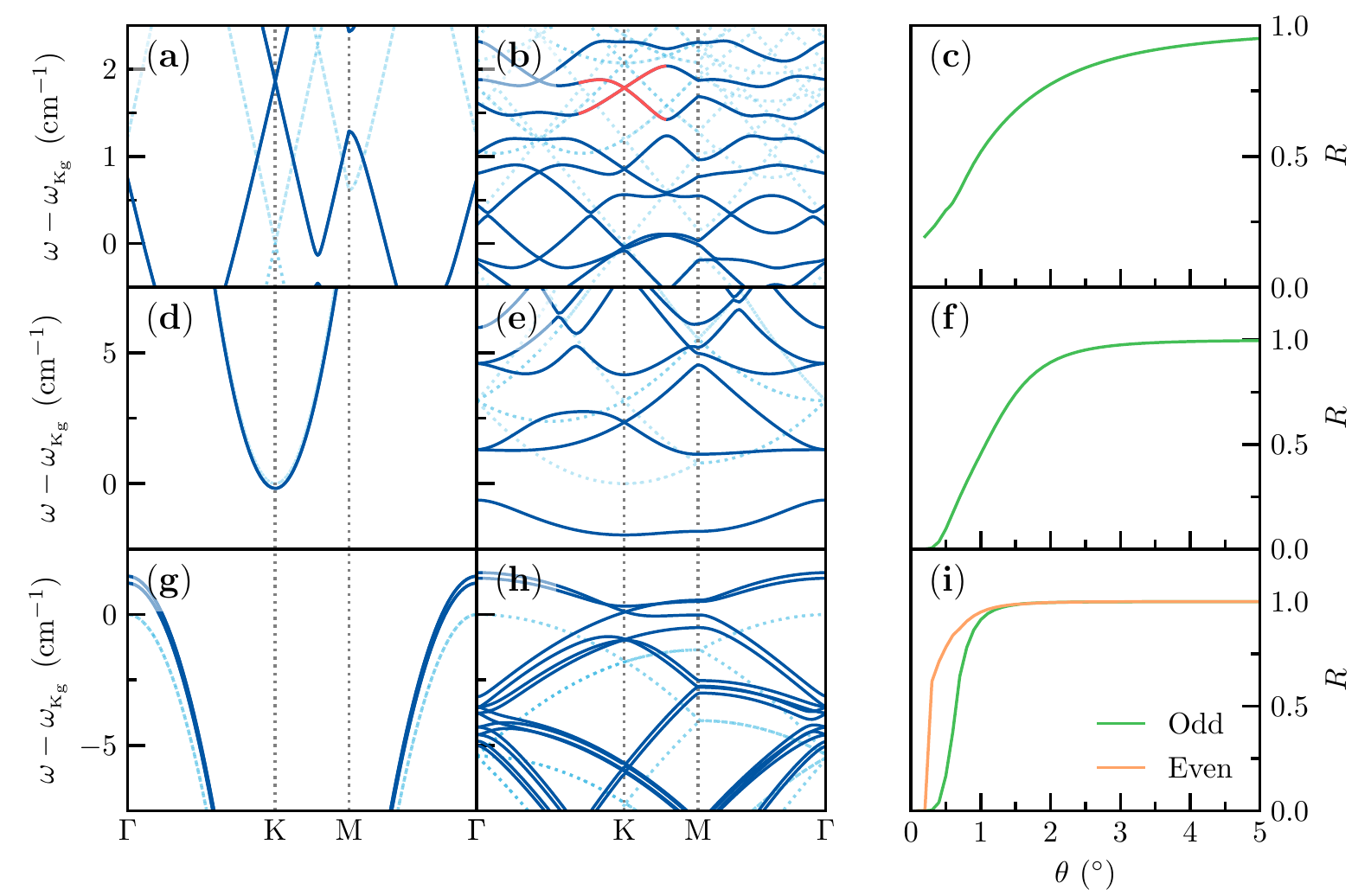}
    \caption{Evolution with the twist angle of the phonon dispersion in the
    moir\'e Brillouin zone for the LO/LA modes at K (top panels),
     the TO mode at K (middle panels), and the TO modes at $\Gamma$
    (bottom panels).
    Left panels are for twist angle $\theta=4^\circ$,
    central panels for $\theta=1.05^\circ$. 
    In panel (b) the relevant Dirac modes in the twisted cases are highlighted in red color.
    In the right panels we show 
    the band renormalization factor $R$ for each mode
     as a function of twist angle.
    }
    \label{fig:bands_twist}
\end{figure*}
{\it Dirac phonons at K.}
The Dirac phonons at the K point are more conveniently described
by introducing a chiral basis
$\tilde{u}_{\alpha,\nu}({\bf q})$, where $\nu=$ R, L
and $\tilde{u}_{\alpha,{\rm R/L}}=(u_{\alpha,x}\pm i u_{\alpha,y})/\sqrt{2}$.
The dynamical matrices for the AA and AB structures in this basis 
read:
\begin{eqnarray}
\hat{\bf K}^{f}({\bf q})
&=&
\sum_\nu
\hat{f}_\nu \hat{\sigma}_0
\left[
 \hat{\tau}_0
 - \pi^\prime_\nu({\bf q}) \hat{\tau}_x
+ \pi^{\prime\prime}_\nu({\bf q})  \hat{\tau}_y
 \right]
,
\\
\hat{\bf K}^{f^\prime}_{\rm AA}({\bf q})
&=&
\hat{f}^\prime_\perp
\left[
 \hat{\sigma}_0
+ \hat{\sigma}_x
\right]
 \hat{\tau}_0
,
\\
\hat{\bf K}^{f^\prime}_{\rm AB}({\bf q})
&=&
\hat{f}^\prime_\perp
\left[
 \hat{\sigma}_0\hat{\tau}_0
- \hat{\sigma}_z\hat{\tau}_z
+\hat{\sigma}_x\hat{\tau}_x
+\hat{\sigma}_y\hat{\tau}_y
\right]/2,
\\
\hat{\bf K}^{f^{\prime\prime}}_{\rm AA}({\bf q})
&=&
\sum_\nu
\hat{f}^{\prime\prime}_\nu
\left[
 \hat{\sigma}_0
 \hat{\tau}_0
-
\pi^\prime_\nu({\bf q})  \hat{\sigma}_x\hat{\tau}_x
+
i
\pi^{\prime\prime}_\nu({\bf q})  
\hat{\sigma}_y\hat{\tau}_y
\right],
\\
\hat{\bf K}^{f^{\prime\prime}}_{\rm AB}({\bf q})
&=&
\sum_\nu
\hat{f}^{\prime\prime}_\nu
\{[3\hat{\sigma}_0
 \hat{\tau}_0
-
\hat{\sigma}_z\hat{\tau}_z
]/2
\\
&&
-\sum_\nu
\hat{f}^{\prime\prime}_\nu
\pi^\prime_\nu({\bf q})  
[
\hat{\sigma}_x\hat{\tau}_0
-(\hat{\sigma}_x\hat{\tau}_x-\hat{\sigma}_x\hat{\tau}_x)/2]
\nonumber\\
&&
+
\sum_\nu
\hat{f}^{\prime\prime}_\nu
\pi^{\prime\prime}_\nu({\bf q})
[ i\hat{\sigma}_y\hat{\tau}_0
-(\hat{\sigma}_x\hat{\tau}_y+\hat{\sigma}_y\hat{\tau}_x)/2
]\}\nonumber
,
\end{eqnarray}
where $\hat{\tau}_i$ are Pauli matrices acting in the (A, B) sublattice space,
$\hat{\sigma}_i$ are Pauli matrices acting in the layer space,
and $\hat{f}_\nu$, $\hat{f}^\prime_\nu$, $\hat{f}^{\prime\prime}_\nu$  are $2 \times 2$ matrices defined in the (R,L) chiral space,
whose explicit expressions are reported in the Supplementary Material (SM)~\cite{SM}.
The index $\nu$ runs over the three vectors of the in-plane nearest
neighbor B atoms with respect to an atom A, determining also
the effective phonon dispersion
by the non-local factors $\pi^\prime_\nu({\bf q})=\mbox{Re}\{
\exp[i {\bf q}\cdot {\bm \delta}_\nu]\}$,
$\pi^{\prime\prime}_\nu({\bf q})=\mbox{Im}\{
\exp[i {\bf q}\cdot {\bm \delta}_\nu]\}$,
${\bm \delta}_1=(1,0)$, ${\bm \delta}_2=(-1/2,\sqrt{3}/2)$, 
${\bm \delta}_3=(-1/2,-\sqrt{3}/2)$.
Note that the term $\hat{\bf K}^{f}({\bf q})$
in the dynamical matrix does not depend on the specific AA or
AB (or twisted) structure since it is purely related to 
intra-layer physics.

Without interlayer coupling, the phonon dispersion exhibits two degenerate Dirac cones at the K point, emerging from the longitudinal acoustic (LA) and longitudinal optical (LO) branches for each layer. In an AA structure, these cones split into two, while only one survives in AB stacking and the other one is gapped due to the interlayer coupling. To determine the intralayer FC parameters 
$f_\parallel$ and $f_\perp$
we fix the energy of the single-layer Dirac point
$\omega_0=\sqrt{3(f_\parallel+f_\perp)/2M}$,
and their Dirac velocity 
$v=\omega_0 a (f_\parallel-f_\perp)/4(f_\parallel+f_\perp)$.
The other interlayer elastic three parameters
$f^\prime_\perp$, $f^{\prime\prime}_\parallel$,
$f^{\prime\prime}_\perp$ can be determined by fixing the energies of the two Dirac cones
in the AA structure, $\omega_{{\rm AA},\pm}$ \cite{noteeigen}
and by the splitting energy of the single-degenerate levels
in the AB stacking $\omega_{{\rm AB},\pm}$ \cite{SM}.
Using first-principles calculations~\cite{Giannozzi2020} (see SM~\cite{SM}), we obtain $f_\parallel=23.882$ meV/\AA$^2$,
$f_\perp=19.973$ meV/\AA$^2$,
$f^\prime_\perp=-0.143$ meV/\AA$^2$,
$f^{\prime\prime}_\parallel = 0.090$ meV/\AA$^2$,
$f^{\prime\prime}_\perp = 0.059$ meV/\AA$^2$.

Utilizing the dynamical matrix of two uncoupled layers and that of the AA and AB structures, we construct a continuum model in the twisted case.
Here we investigate the effects of twist on the properties
of few selected in-plane lattice modes,
namely the Dirac phonons at the K point
emerging from the LA
and LO branches,
the non-degenerate high-frequency TO mode at the K point. Furthermore, we study the degenerate LO and TO modes at the $\Gamma$ point.
For Dirac phonons,
we restrict the analysis to the relevant 4-fold Hilbert sub-space
containing
the left-hand chiral displacements for the A1/A2 atoms,
and the right-hand chiral displacements for the B1/B2 atoms.
The $4 \times 4$ dynamical matrices so obtained read:

\begin{eqnarray}
\hat{\cal K}_{\rm AA}(\tilde{{\bf q}})
&=&
v \hat{\sigma}_0 [\tilde{q}_x \hat{\tau}_x+\tilde{q}_y \hat{\tau}_y]
+
V_{\rm AA} \hat{\sigma}_0  \hat{\tau}_0
-
f^\prime_\perp \hat{\sigma}_x  \hat{\tau}_0
,
\label{reducedAA}
\\
\hat{\cal K}_{\rm AB}(\tilde{{\bf q}})
&=&
v \hat{\sigma}_0 [\tilde{q}_x \hat{\tau}_x+\tilde{q}_y \hat{\tau}_y]
+
V_{\rm AB,0} \hat{\sigma}_0  \hat{\tau}_0
+
V_{{\rm AB},z} \hat{\sigma}_z  \hat{\tau}_z
\nonumber\\
&&
-
3(f^{\prime\prime}_\parallel-f^{\prime\prime}_\perp)/4
[ \hat{\sigma}_x  \hat{\tau}_x +\hat{\sigma}_y  \hat{\tau}_y]
,
\label{reducedAB}
\end{eqnarray}
where $\tilde{q}_i$ are wave-vectors measured with respect to the K point and
where the parameters $V_{\rm AA}$, $V_{\rm AB,0}$, $V_{\rm AB,z}$
are ruled by the interlayer force constants (for an explicit expression see the SM~\cite{SM}).

Eqs. (\ref{reducedAA})-(\ref{reducedAB}) provide
the basis for assessing the evolution of the Dirac phonons
in TBG within a continuum model.
Following a similar approach as for electrons,
we describe the dynamical matrix for a twisted bilayer by interpolating the 
off-diagonal blocks of the AA and AB matrices in Eqs. (\ref{reducedAA})-(\ref{reducedAB}) \cite{LPN07}.
We find that the equivalent of an AA and AB interlayer tunneling
are ruled by the terms:
\begin{align}
    t_{\rm AA} &= - f'_\perp/3, \\
    t_{\rm AB} &= - (f''_\parallel - f''_\perp )/2 .
    \label{hoppings}
\end{align}
Furthermore, we  notice that the diagonal elements 
of Eqs. (\ref{reducedAA})-(\ref{reducedAB}) give rise to effective
local potentials which are different for different local stackings,
and hence in different regions in real space corresponding to an AA, AB or BA stacking \cite{SM}.
These potentials can be expanded in reciprocal lattice vectors similarly to the way electrostatic potentials
are incorporated into the continuum model for electronic bands of TBG~\cite{GW18}.
Including in this scheme these local potentials,
we show in Fig. \ref{fig:bands_twist}a,b the evolution with the twist angle of the phonon dispersion close to the
in-plane Dirac energies in the moir\'e Brillouin zone.
We can notice an overall upward energy shift of all the
phonon frequencies, stemming from the presence of
such local potentials.
Moreover, the phonon dispersion still shows a dispersive
Dirac behavior close to K for $\theta=4^\circ$,
with a linear dispersion velocity comparable with the single-layer.
Interestingly, such dispersion appears much flatter at $\theta=1.05^\circ$,
signalizing a remarkable band renormalization.
In order to get a qualitative estimate of a possible ``phonon magic angle'',
we can employ the standard approach of truncating the interlayer
tunneling only to the first set of three momenta
${\bf Q}_\nu$  \cite{BM11} and we get $\bar{\theta}_{\rm LO/LA}
\approx 2.1^\circ$ (see SM~\cite{SM}).
Such a picture is supported by
a quantitative analysis based on multi-interlayer scattering,
including local potentials.
Within this framework, following Ref. \cite{BM11},
the flattening of the LO/LA phonon bands
can be parameterized in terms of the renormalization factor $R=V^*/V$
of the Dirac phonon velocity $V^*$ in the twisted case with respect to the one in the single-layer, $V$.
The twist-angle dependence of $R$ for the full
multi-scattering continuum model
is plotted in Fig.~\ref{fig:bands_twist}c, showing a marked
depletion for twist angles $\lesssim 2^\circ$.
Nevertheless, such depletion, for these as for other lattice modes,
never reach a perfect flattening because of the role of the local potentials,
the qualitative estimate of the phonon magic angle
can properly capture the correct range of twist angles where
a strong phonon band renormalization occurs.

{\it TO phonon at K.}
The analysis done for Dirac phonons at K can be also extended to the TO phonon at K, which induces intervalley scattering for the electrons \cite{ATF19}.
Following the usual scheme,
the phonon wavefunction is expanded in plane waves in the two layers,
where the plane wave in one layer is transferred as a superposition of three plane waves in the neighboring layer (see SM~\cite{SM} for more details).
The main differences with respect to the LO/LA modes are: $i$) The monolayer TO phonon is not degenerate at K,
so that the spinor (sublattice) degree of freedom disappears; 
$ii$) The dispersion in the monolayer is quadratic in $\tilde{\bf q}$.
The coupling between layers includes a diagonal restoring term, which changes from the AA to the AB and BA regions, and a single interlayer coupling, which also depends on the position within the unit cell. This coupling is finite in the AA region, and it vanishes in the AB and BA regions \cite{SM}. Hence, the model includes four parameters, which can be readily obtained from the FCs discussed above. The model used here resembles the ones used for the conduction band edge of MoS$_2$ (located at the K and K$^\prime$ points)\cite{WLTMM19,CFK20}.
The representative plots of the TO phonon dispersion
in the moir\'e Brillouin zone 
are shown in Fig.~\ref{fig:bands_twist}d,e,
and the angle dependence of the appropriate band renormalization
for the TO modes is depicted in Fig.~\ref{fig:bands_twist}f.
Using the standard approximate model restricted to the first star of Bloch waves and neglecting the diagonal restoring forces,
we can obtain also for these modes an estimate for the magic
angle at which the prefactor of the quadratic dispersion at K
vanishes~\cite{SM}. We obtain $\bar{\theta}_{TO} \approx 1.0^\circ$, which is qualitatively
consistent with the results shown in Fig.~\ref{fig:bands_twist}f.

{\it Optical phonons at $\Gamma$.}
The continuum model for the optical phonons at $\Gamma$ in TBG
is particularly simplified by the fact that each plane-wave
in one layer just tunnels into a single plane-wave in the other layer.
As detailed in the SM \cite{SM}, the interlayer forces thus couple separately
the LO and the TO modes.
One can further divide modes with even and odd symmetry 
with respect to the vertical axis.
The LO and TO modes of the single-layer evolve thus
in TBG into four {\sl independent} bands with a quadratic dispersion
which is ruled by different combinations of the FC parameters,
and hence with four different
behaviors for the band renormalization \cite{SM}.
The model resembles electronic models used for the valence band edge of MoS$_2$ (located at the $\Gamma$ point)\cite{ZLF21,WG21,AM21}.
The plots of the phonon dispersion of the TO modes
with even and odd symmetry for different twist angles
is shown in Fig.~\ref{fig:bands_twist}g,h,
and the angle dependence of the effective band renormalization
in Fig.~\ref{fig:bands_twist}i.
Similar results (not shown) are obtained for the LO modes.

{\it Discussion.}
We have analyzed the optical phonons of TBG, by introducing proper continuum models originally devised for the electronic structure.
For all the three cases studied, LO/LA modes at K, TO modes at K and LO/TO modes at $\Gamma$, we find a remarkable flattening of the superlattice phonon bands at low twist angles, starting at higher values than the ``magic angles" where electronic flat bands appear. 
The onset of such flat phonon bands is expected to 
tune the optical properties of TBG in the infrared frequency range, providing a possible tool for twist characterization.
LO/TO modes are directly probed by one-phonon Raman and infrared spectroscopy
in bilayer graphene \cite{wang09}, with intensities and selection rules
that depend crucially on the bilayer stacking order and on
the $z$-axis symmetry \cite{malard08,gava09,kuzmenko09,tang09,cappelluti12}, and hence on twisting \cite{jorio14,popov18,GR20,chang22}.
TO modes, and their dispersions close to the K point,
are also commonly observed by means of double-resonance
processes D and 2D \cite{ferrari00,Petal04,Fetal06,basko07,B08}.
Finally, although a direct contribution of the
LO/LA modes at K to the Raman phonon spectroscopy is not well-assessed \cite{matthews99,thomsen00,ferrari00,Fetal06}, these modes bare
a promising relevance for quantum devices since,
obeying to a similar Dirac quantum-structure, they are
expected to show a similar rich complexity
as the electronic degree of freedom.
It is also worth mentioning that the same modes
in the presence of mass disproportion (e.g. in h-BN)
host chiral phonon states supporting a finite lattice angular
momentum \cite{zhang14,zhang15,zhu2018,chen2019},
with possible application towards a suitable (lattice-based) quantum
two-level systems \cite{suri21,maity22}.

{\it Acknowledgments}
All the authors thank T. Cea for useful discussions.
IMDEA Nanociencia acknowledges support from the ``Severo Ochoa" Programme for Centres of Excellence in R\&D (CEX2020-001039-S / AEI / 10.13039/501100011033).
F.G. acknowledges funding from the European Commission, within the Graphene Flagship, Core 3, grant number 881603 and from grants NMAT2D (Comunidad de Madrid, Spain), SprQuMat (Ministerio de Ciencia e Innovaci\'on, Spain) and financial support through the (MAD2D-CM)-MRR MATERIALES AVANZADOS-IMDEA-NC. 
E.C. acknowledges financial support from PNRR MUR project PE0000023-NQSTI. H.R. acknowledges the support from the Swedish Research Council (VR Starting Grant No. 2018-04252).


\clearpage
\newpage


\setcounter{equation}{0}
\setcounter{figure}{0}
\setcounter{table}{0}
\setcounter{page}{1}
\setcounter{section}{0}
\makeatletter
\renewcommand{\thesection}{\Alph{section}}
\renewcommand{\theequation}{S\arabic{equation}}
\renewcommand{\thefigure}{S\arabic{figure}}
\renewcommand{\thetable}{S\arabic{table}}
\renewcommand{\thepage}{SM-\arabic{page}}

\onecolumngrid
\begin{center}
\textbf{\large Supplemental Material for:\\Flat-band optical phonons in twisted bilayer graphene}
\end{center}
\section{Force-constant model}

In this Section,
we provide details about the force-constant model employed in the present paper
to describe the lattice dynamics in monolayer, bilayers and twisted bilayer graphenes.
We focus here only on in-plane lattice displacements that, due to their
mixing of $x$ and $y$ component, show the most interesting physics.
A similar model can be employed for out-of-plane lattice displacements.

Building blocks of such model are forces between nearest neighbor pairs of atoms connected by a vector ${\bf r}_{ij}$.
Two main components can be thus identified, a parallel one with respect to ${\bf r}_{ij}$ (central forces);
and a perpendicular one to this vector (transverse forces) and lying in the graphene plane.
A third component, orthogonal to the other two ones, can be also included,
but it mainly rules the out-of-plane lattice dynamics and it does not play
any relevant role in the present context.

\subsection{Single-layer graphene}

For an isolated graphene layer
the dynamical matrix  is defined in terms of the atomic lattice displacements, $\{ x_A , y_A ,x_B , y_B \}$, were
$A$ , $B$ are the sublattice labels.
Following the above notation, we introduce the central and transverse  forces
governed by the parameter $f_\parallel$, $f_\perp$ respectively.
The resulting $4 \times 4$ matrix reads thus:
\begin{align}
    {H}(\vec{k}) &
    = 
    \left( \begin{array}{cc} 
{H}_{AA}(\vec{k}) & {H}_{AB}(\vec{k}) \\
{H}_{AB}^\dagger(\vec{k}) &  {H}_{BB}(\vec{k})
  \end{array} \right) 
    ,
    \label{hamil1}
\end{align} 
where
\begin{align}
    {H}_{AA}(\vec{k}) &
    = 
    {H}_{BB} (\vec{k})
    = 
    \left( \begin{array}{cc} 
    3 \left[ f_\parallel + f_\perp \right]/2    & 0 \\
    0 
    & 3 \left[ f_\parallel + f_\perp \right]/2
  \end{array} \right) 
    ,
    \label{hamil1a}
\end{align} 
and
\begin{align}
    {H}_{AB}(\vec{k}) &
    = 
    \left( \begin{array}{cc} 
3 f_\parallel \left[ e_{\vec{k},a} + e_{\vec{k},b} \right]/4 + f_\perp \left[ 1 + (e_{\vec{k},a} + e_{\vec{k},b})/4 \right] 
    & - \sqrt{3} \left[ f_\parallel - f_\perp \right] \left( e_{\vec{k},a} - e_{\vec{k},b} \right)/4
    \\
    \sqrt{3} \left[ f_\parallel - f_\perp \right] \left( e_{\vec{k},a} - e_{\vec{k},b} \right)/4
    & f_\parallel \left[ 1 + (e_{\vec{k},a} + e_{\vec{k},b})/4 \right] + 3 f_\perp \left( e_{\vec{k},a} + e_{\vec{k},b} \right)/4
  \end{array} \right) 
    .
    \label{hamil1b}
\end{align} 

Here $e_{\vec{k},i}=\exp[i \vec{k} \cdot \vec{R}_i]$ ($i=a,b$), where
$\vec{R}_a=a ( 1/2 , - \sqrt{3}/2 )$, $\vec{R}_b=a ( -1/2 , \sqrt{3}/2 )$, 
and $a \approx 2.46$ \AA \, is the lattice constant.

The model in Eqs. (\ref{hamil1})-(\ref{hamil1b}) has simple solutions at high symmetry points:
\begin{align}
  M \omega^2_\alpha(\Gamma) &= \left\{ 0 , 0 , 3 \left( f_\parallel  + f_\perp \right) , 3 \left( f_\parallel  + f_\perp \right) \right\}, \nonumber \\
   M \omega^2_\alpha(K) &= \left\{ 3 f_\parallel , 3 f_\perp  ,  \frac{3 \left( f_\parallel  + f_\perp \right)}{2} , \frac{3 \left( f_\parallel  + f_\perp \right)}{2} \right \} ,
  \nonumber \\
   M \omega^2_\alpha(M) &= \left\{ 2 f_\parallel , 2 f_\perp  ,  3 f_\parallel + f_\perp , f_\parallel  + 3 f_\perp  \right \} ,
\end{align}
where the index $\alpha$ labels the phonon branch
and the parameter $M$ is the mass of the carbon atom. 

We focus on characteristic modes at the high-symmetry points
$\Gamma$, K, relevant for twisted system. 

The first ones are the the double-degenerate
high-frequency states at the $\Gamma$ point, 
which represent the longitudinal optical (LO) and the transverse optical (TO) modes.
These modes determine the full phonon bandwidth,
given by the energy
\begin{eqnarray}
M \omega^2_{LO/TO}(\Gamma)
&=&
3 \left( f_\parallel  + f_\perp \right)
.
\end{eqnarray}
The corresponding eigenvectors for these modes are:
\begin{align}
{\bm \epsilon}_{LO/TO,x}(\Gamma)
&
=
\frac{1}{\sqrt{2}}
 \left( \begin{array}{c} 1 \\ 0 \\ -1 \\ 0 \end{array} \right) ,
\hspace{1cm}
{\bm \epsilon}_{LO/TO,y}(\Gamma)
=
\frac{1}{\sqrt{2}} \left( \begin{array}{c} 0 \\ 1 \\0 \\ - 1 \end{array} \right) 
.
\label{eigenTOLO}
\end{align}

Quite relevant is also the evolution of the TO branch
at the K point, which leads to a
the high-frequency mode at the K point, with energy 
\begin{eqnarray}
M \omega^2_{TO}(K)=3 f_\parallel 
,
\end{eqnarray}
and a typical eigenvector for
transverse-optical displacements:
\begin{align}
{\bm \epsilon}_{TO}(K)
&
=
\frac{1}{2}
 \left( \begin{array}{c} 1 \\ -i \\ -1 \\ -i \end{array} \right) .
\label{etok}
\end{align}

The spectrum at the K point is further characterized by the doublet with energy
\begin{eqnarray}
M \omega^2_{LO/LA}(K)
&=&
3 \left( f_\parallel  + f_\perp \right)/2
.
\label{LOLA_1L}
\end{eqnarray}
The eigenstates for these modes at the K point can be written as:
\begin{align}
{\bm \epsilon}_{LO/LA,+}(K)
&
=
\frac{1}{\sqrt{2}}
 \left( \begin{array}{c} 1 \\ i \\ 0 \\ 0 \end{array} \right) ,
\hspace{1cm}
{\bm \epsilon}_{LO/LA,-}(K)
=
\frac{1}{\sqrt{2}} \left( \begin{array}{c} 0 \\ 0 \\1 \\ - i \end{array} \right) 
    \label{wvk}
\end{align}
The eigenstates in the K$^\prime$ point can be obtained by reversing the sign
in front of the imaginary terms.
Using such eigenstates as reduced Hilbert space, and a ${\bf k} \cdot {\bf p}$ expansion,
the dynamical matrix ${\cal \tilde{H}}$ restricted to the
closeness of the $K$ and $K'$ points can be thus approximated as:
\begin{align}
  {\cal H}_{LO/LA}(\vec{k}) &=  \left( 
    \begin{array}{cc} 
    3 \left( f_\parallel  + f_\perp \right)/2
    & V a \left( \xi k_x + i k_y \right) \\
    V a \left( \xi k_x - i k_y \right) 
    & 3 \left( f_\parallel  + f_\perp \right)/2
    \end{array}
    \right)
    \label{hamilk}
\end{align}
where $\xi = \pm 1$ for the K, K$^\prime$ respectively.
The phonon dispersion for these modes close to the
K point can be written thus as:
\begin{eqnarray}
E^\pm_{LO/LA}(\vec{k})
&=&
M \omega^2_{LO/LA}(K)
=
\frac{3}{2} \left( f_\parallel  + f_\perp \right)
\pm
Va |\vec{k}|
.
\label{dynDirac}
\end{eqnarray}
The term $V= \sqrt{3} | f_\parallel - f_\perp | /4$ rules here the linear Dirac dependence of the {\em dynamical} matrix
close to the K/K$^\prime$ point, not to be confused with the linear slope of the phonon dispersion.

\subsection{Graphene bilayers.}

The force-constant model described above for single-layer graphene can be extended
in a compelling way to graphene bilayers.
We analyze in details in the following
the AA and the AB stacking structures,
whereas the model for BA stacking can be obtained
from the AB by switching the sublattice space indeces.
We include interlayer force constants only up to the second nearest neighbors carbon pairs.
In both cases interlayer nearest neighbors are represented by carbon atoms lying directly on top of each other,
at the interlayer distance, $d \approx 3.4$\AA.
The only elastic term between these atoms is the transverse one, ruled by $f'_\perp$.
Second interlayer nearest neighbors are represented by atoms at the interatomic distance $\sqrt{d^2+a^2/3} \approx 3.5$\AA.
In the case both the two parallel and transverse components need to be taken into account,
parametrized by the terms $f''_\parallel , f''_\perp$.

Using the 8-fold spinor represented by the lattice displacements for both layers,
for a generic bilayer graphene system,
the dynamical matrix can be written as:
\begin{align}
    {H}^{2L}(\vec{k}) &
=
    {H}^{1L+1L}(\vec{k})
+
\delta {H}(\vec{k})
,    \end{align} 
where
${H}^{1L+1L}(\vec{k})$ is the dynamical matrix for the
two decoupled layers and $\delta {H}(\vec{k})$ takes into account
the interlayer forces.
It is clear that $8 \times 8$ ${H}^{1L+1L}(\vec{k})$ can be written
as a block-diagonal matrix,
\begin{align}
    {H}^{1L+1L}(\vec{k})
&
=
 \left( \begin{array}{cc} 
{H}(\vec{k}) & 0 \\
0 & {H}(\vec{k})
  \end{array} \right) 
, 
 \end{align} 
whereas
\begin{align}
  \delta{H}(\vec{k})  &
=
 \left( \begin{array}{cc} 
\delta{H}_{11}(\vec{k})  &
\delta{H}_{12}(\vec{k})  \\
\delta{H}_{12}^\dagger(\vec{k})   & \delta{H}_{22}(\vec{k}) 
  \end{array} \right) 
, 
 \end{align} 
which contains
both block on-diagonal terms, resulting from
the quadratic elastic contributions associated 
with a single-atom lattice displacements,
and block off-diagonal terms which describe the effective
interlayer elastic forces.

More explicitely, for the AA and AB structures we have respectively:
\begin{align}
    \delta {H}_{11}^{AA} (\vec{k})&
    =   \delta {H}_{22}^{AA}(\vec{k}) = 
    \left( \begin{array}{cccc} 
    f'_\perp + \frac{\displaystyle 3 \left( f''_\parallel + f''_\perp \right)}{2} 
    &0 
    &0 
    &0 \\ 
    0 
    & f'_\perp + \frac{\displaystyle 3 \left( f''_\parallel + f''_\perp \right)}{\displaystyle 2}
    &0 
    &0 \\ 
    0 
    &0 
    &f'_\perp + \frac{\displaystyle 3 \left( f''_\parallel + f''_\perp \right)}{\displaystyle 2} 
    &0 \\ 
    0 
    &0 
    &0 
    &f'_\perp + \frac{\displaystyle 3 \left( f''_\parallel + f''_\perp \right)}{\displaystyle 2}
    \end{array} \right)
    ,
\end{align}
and
\begin{align}
     \delta {H}_{11}^{AB}(\vec{k}) &= \left( \begin{array}{cccc} 
     f'_\perp + \frac{\displaystyle 3 \left( f''_\parallel + f''_\perp \right)}{\displaystyle 2} 
     &0 
     &0
     &0 \\ 
     0 
     & f'_\perp + \frac{\displaystyle 3 \left( f''_\parallel + f''_\perp \right)}{\displaystyle 2}
     &0 
     &0 \\ 
     0 
     &0  
     &3 \left( f''_\parallel + f''_\perp \right) 
     &0 \\ 
     0 
     &0 
     &0 
     & 3 \left( f''_\parallel + f''_\perp \right)\end{array} \right)
     \nonumber \\
     \delta {H}_{\vec{k}}^{AB,22} &=\left( \begin{array}{cccc} 
     3 \left( f''_\parallel + f''_\perp \right) 
     &0 
     &0 
     &0 \\ 
     0 
     &  3 \left( f''_\parallel + f''_\perp \right)
     &0 
     &0 \\ 
     0 
     &0 
     &f'_\perp + \frac{\displaystyle 3 \left( f''_\parallel + f''_\perp \right)}{\displaystyle 2} 
     &0 \\ 
     0 
     &0 
     &0 
     &f'_\perp + \frac{\displaystyle 3 \left( f''_\parallel + f''_\perp \right)}{\displaystyle 2}\end{array} \right)
\end{align}


In similar way,
the interlayer forces for a generic bilayer structures read:
\begin{align}
    \delta{H}_{12}^{\alpha}(\vec{k}) &
    = 
    \left( \begin{array}{cc} 
\delta{H}_{12,AA}^{\alpha}(\vec{k}) & \delta{H}_{12,AB}^{\alpha}(\vec{k}) \\
\delta{H}_{12,BA}^{\alpha}(\vec{k})&  \delta{H}_{12,BB}^{\alpha}(\vec{k})
  \end{array} \right) 
    .
\end{align}

Note that here the upper label $\alpha=AA,AB$ denotes the {\em stacking} order,
whereas the indeces  AA, AB in the subscript represent the {\em sublattice} label for each layer.
Using these notations, we obtain:
\begin{align}
\delta{H}_{12,AA}^{AA}(\vec{k})
    & = 
    \delta{H}_{12,BB}^{AA}(\vec{k})
    =
    \left( \begin{array}{cc} 
     - f'_\perp    & 0 \\
     0 
  &- f'_\perp 
  \end{array} \right) 
   ,
   \end{align}
\begin{eqnarray}
\delta{H}_{12,AB}^{AA}(\vec{k})
    & = &
   \delta{H}_{12,AB}^{AA}(\vec{k})
   \nonumber\\
    &=&
    \left( \begin{array}{cc} 
 -\frac{\displaystyle 3}{\displaystyle 4} f''_\parallel \left( e_{\vec{k},a} + e_{\vec{k},b} \right) 
 + \frac{\displaystyle f''_\perp}{\displaystyle 4} \left[ 1 + (e_{\vec{k},a} + e_{\vec{k},b}) \right] 
  & \frac{\displaystyle \sqrt{3}}{\displaystyle 4} \left[ f''_\parallel - f''_\perp \right] \left( e_{\vec{k},a} - e_{\vec{k},b} \right) \\
  \frac{\displaystyle \sqrt{3}}{\displaystyle 4} \left[ f''_\parallel - f''_\perp \right] \left( e_{\vec{k},a} - e_{\vec{k},b} \right)
   & \frac{\displaystyle 3}{\displaystyle 4} f''_\parallel \left( e_{\vec{k},a} + e_{\vec{k},b} \right)
   + \frac{\displaystyle f''_\perp}{\displaystyle 4} \left[ 1 + (e_{\vec{k},a} + e_{\vec{k},b}) \right]
  \end{array} \right) 
   ,
   \end{eqnarray}
\begin{eqnarray}
\delta{H}_{12,AA}^{AB}(\vec{k})
    & = &
    \delta{H}_{12,BB}^{AB}(\vec{k})
\nonumber\\
    &=&
    \left( \begin{array}{cc} 
- \frac{\displaystyle 3}{\displaystyle 4} f''_\parallel \left( e_{\vec{k},a}^* + e_{\vec{k},b}^* \right) 
- \frac{\displaystyle f''_\perp}{\displaystyle 4} \left[ 1 + (e_{\vec{k},a}^* + e_{\vec{k},b}^*) \right] 
& \frac{\displaystyle \sqrt{3}}{\displaystyle 4} \left[ f''_\parallel - f''_\perp \right] \left( - e_{\vec{k},a}^* + e_{\vec{k},b}^* \right)
\\
\frac{\displaystyle \sqrt{3}}{\displaystyle 4} \left[ f''_\parallel - f''_\perp \right] \left( - e_{\vec{k},a}^* + e_{\vec{k},b}^* \right)
& - \frac{\displaystyle 3}{\displaystyle 4} f''_\parallel \left( e_{\vec{k},a}^* + e_{\vec{k},b}^* \right) - \frac{\displaystyle f''_\perp}{\displaystyle 4} \left[ 1 + (e_{\vec{k},a}^* + e_{\vec{k},b}^*) \right]
  \end{array} \right) 
   ,
   \end{eqnarray}
\begin{eqnarray}
\delta{H}_{12,AB}^{AB}(\vec{k})
    & = &
    \left( \begin{array}{cc} 
- f'_\perp &0 \\
0
&- f'_\perp
  \end{array} \right) 
   ,
   \end{eqnarray}
\begin{eqnarray}
\delta{H}_{12,BA}^{AB}(\vec{k})
    & = &
    \left( \begin{array}{cc} 
- \frac{\displaystyle 3}{\displaystyle 4} f''_\parallel
( e_{\vec{k},a}^* + e_{\vec{k},b}^* )
- \frac{\displaystyle f''_\perp}{\displaystyle 4} \left[ e_{\vec{k},a}^* e_{\vec{k},b}^* + (e_{\vec{k},a}^* + e_{\vec{k},b}^*) \right]
&-  \frac{\displaystyle\sqrt{3}}{\displaystyle 4} [ f''_\parallel - f''_\perp ] ( e_{\vec{k},a}^* - e_{\vec{k},b}^*)
\\
- \frac{\displaystyle \sqrt{3}}{\displaystyle 4} [ f''_\parallel - f''_\perp ] ( e_{\vec{k},a}^* - e_{\vec{k},b}^* )
  &- \frac{\displaystyle 3}{\displaystyle 4} f''_\parallel ( e_{\vec{k},a}^* + e_{\vec{k},b}^* )
  - \frac{\displaystyle f''_\perp}{\displaystyle 4} \left[ e_{\vec{k},a}^* e_{\vec{k},b}^* + (e_{\vec{k},a}^* + e_{\vec{k},b}^*)\right]
  \end{array} \right) 
   ,
   \end{eqnarray}

On the ground of the knowledge of the full dynamical matrix for generic branch and generic momentum,
we can now build up the effective model
for selected modes in the bilayer structures,
in the spirit of a ${\bf k}\cdot {\bf p}$
expansion.

\subsubsection{Dirac LO/LA modes at K point}

We first address the LO/LA modes at the K point,
which gives rise to a linear Dirac-like phonon dispersion.

Using the eigenstates (\ref{wvk}) in each layer,
we can write the effective dynamical matrix
for a bilayer with generic structure $\alpha=AA, AB$ in
a reduced $4 \times 4$ Hilbert space:
\begin{align}
    {{\cal H}}_{LO/LA}^{\alpha}(\vec{k}) &= \left(
    \begin{array}{cc}
     {{\cal H}}_{LO/LA}(\vec{k})+ {\cal V}_{LO/LA,1}^{\alpha} &
     {{\cal H}}_{LO/LA,12}^{\alpha} \\
    {{\cal H}}_{LO/LA,12}^{\alpha,\dagger} & 
    {{\cal H}}_{LO/LA}(\vec{k})
    + {\cal V}_{LO/LA,2}^{\alpha}
\end{array}
\right),
\end{align}
where ${\cal H}_{LO/LA}(\vec{k})$ is the 
${\bf k}\cdot {\bf p}$ expansion for the LO/LA modes
at the K point for a single-layer, as reported in
Eq. (\ref{hamilk}), ${{\cal H}}_{LO/LA,12}^{\alpha}$ is the
$2 \times 2$ interlayer coupling matrix which is different
for different stacking orders, namely:
\begin{align}
  {\cal H}_{LO/LA,12}^{AA}  &= \left(
    \begin{array}{cc}
- f'_\perp & 0\\
0 & - f'_\perp
\end{array}
\right),
\label{interTOTA_AA}
\end{align}
\begin{align}
   \tilde{{\cal H}}_{LO/LA,12}^{AB}&= \left(
    \begin{array}{cc}
0 & 0 \\
-\frac{\displaystyle 3}{\displaystyle 2}(f''_\parallel - f''_\perp ) & 0
\end{array}
\right),
\label{interTOTA_AB}
\end{align}
and where ${\cal V}_{LO/LA,i}^{\alpha}$
are $2 \times 2$ {\em diagonal} matrices
that represent the stacking-dependent
{\em onsite} intra-atomic potential on each layer $i=1,2$
due to interlayer elastic coupling,
explicitely:
\begin{align}
   {\cal V}_{LO/LA,1}^{AA}
   &=
   {\cal V}_{LO/LA,2}^{AA}
   =
   \left(
    \begin{array}{cc}
f'_\perp + \frac{\displaystyle 3}{\displaystyle 2} ( f''_\parallel + f''_\perp )   & 0\\
0 & f'_\perp + \frac{\displaystyle 3}{\displaystyle 2} ( f''_\parallel + f''_\perp )
\end{array}
\right),
\end{align}
\begin{align}
   {\cal V}_{LO/LA,1}^{AB}
   &= \left(
    \begin{array}{cc}
    \frac{\displaystyle 3}{\displaystyle 2} ( f''_\parallel + f''_\perp ) + f'_\perp 
      & 0\\
0 & 3(f''_\parallel + f''_\perp )
\end{array}
\right),
\end{align}
\begin{align}
   {\cal V}_{LO/LA,2}^{AB}
   &= \left(
    \begin{array}{cc}
    3(f''_\parallel + f''_\perp ) & 0 \\
    0 & \frac{\displaystyle 3}{\displaystyle 2} ( f''_\parallel + f''_\perp ) + f'_\perp 
\end{array}
\right).
\end{align}

The corresponding frequencies at the K point
for each stacking structure read thus:
\begin{eqnarray}
    M \omega^2_{{LO/LA},AA}(K)
    &= &
    \left\{ \begin{array}{cc} 
    \frac{\displaystyle 3}{\displaystyle 2} ( f_\parallel + f_\perp ) 
    + 2 f'_\perp 
    + \frac{\displaystyle 3}{\displaystyle 2} ( f''_\parallel + f''_\perp ) & \hspace{5mm} \mbox{(double degenerate)},
    \\ \frac{\displaystyle 3}{\displaystyle 2} ( f_\parallel + f_\perp )  + \frac{\displaystyle 3}{\displaystyle 2} ( f''_\parallel + f''_\perp ) & \hspace{5mm} \mbox{(double 
    degenerate)},
        \end{array}
    \right.
    \label{LOLAsplitAA}
    \\
       M \omega^2_{{LO/LA},AB}(K)
    &= &
    \left\{ \begin{array}{cc} 
    \frac{\displaystyle 3}{\displaystyle 2}( f_\parallel + f_\perp ) + f'_\perp 
    + \frac{\displaystyle 3}{\displaystyle 2} ( f''_\parallel + f''_\perp ) &
    \hspace{5mm} \mbox{(double degenerate)},
    \\
    \frac{\displaystyle 3}{\displaystyle 2} ( f_\parallel + f_\perp )  + \frac{\displaystyle 3}{\displaystyle 2} ( 3 f''_\parallel + f''_\perp ) ,&
     \\ \frac{\displaystyle 3}{\displaystyle 2} ( f_\parallel + f_\perp )  + \frac{\displaystyle 3}{2\displaystyle } ( f''_\parallel + 3 f''_\perp ), &
    \end{array}
 \right.
     \label{LOLAsplitAB}
\end{eqnarray}
Note that, in the AA stacking,
the first state corresponds to out-of-phase lattice
displacements in the two layers, whereas the second
frequency described in-phase in-plane 
lattice vibrations.

In similar way as for the electronic structure,
the interlayer coupling gives rise thus to different effects
according the to the different stacking.
More in particular,
in strict similarity with the
electronic dispersion,
the Dirac-like LO/LA modes
at the K point of the single-layer
gives rise in the AA bilayer to {\em two} Dirac phonon cones
split by the interlayer coupling; while
in the AB structure 
just one Dirac phonon cone survives (check)
whereas the other one is effectively gapped.

\subsubsection{High-energy TO modes at K point}

An effective ${\bf k}\cdot {\bf p}$ model can be built
also for the high-energy TO mode at the K point.
Starting point is this case will be 
the eigenstate ${\bm \epsilon}_{TO}(K)$
as expressed in Eq. (\ref{etok}).
Using such state in each layer,
we can build up a $2 \times 2$
${\bf k}\cdot {\bf p}$ model
close to the K point for these modes
in the bilayer systems as:
\begin{align}
    {{\cal H}}_{TO,K}^{\alpha}(\vec{k}) &= \left(
    \begin{array}{cc}
     {{\cal H}}_{TO,K}(\vec{k})+ {\cal V}_{TO,K}^{\alpha} &
     {{\cal H}}_{TO,K}^{\alpha} \\
    {{\cal H}}_{TO,K}^{\alpha} & 
    {{\cal H}}_{TO,K}(\vec{k})
    + {\cal V}_{TO,K}^{\alpha}
\end{array}
\right),
\end{align}
where ${{\cal H}}_{TO,K}(\vec{k})$
is the dynamical matrix at the quadratic order
of this mode in the single-layer,
\begin{align}
{{\cal H}}_{TO,K}(\vec{k})
&=
3 f_\parallel + \frac{ f_\parallel f_\perp}{2 ( f_\parallel - f_\perp )} ( k_x^2 + k_y^2 ) a^2,
\label{disp_TO}
\end{align}
${{\cal H}}_{TO,K}^{\alpha}$ is the inter-layer coupling,
\begin{align}
{{\cal H}}_{TO,K}^{AA}
&=
- f'_\perp + \frac{3 ( f''_\parallel - f''_\perp )}{2} 
,
\label{interTOK_AA}
\\
{{\cal H}}_{TO,K}^{AB}
&=
0,
\label{interTOK_AB}
\end{align}
and 
${\cal V}_{TO,K}^{\alpha}$
represents the 
{\em onsite} intra-atomic potential:
\begin{align}
{\cal V}_{TO,K}^{AA}
&=
f'_\perp + \frac{3 ( f''_\parallel + f''_\perp )}{2}
,
\\
{\cal V}_{TO,K}^{AB}
&=
\frac{f'_\perp}{2} + \frac{9 ( f''_\parallel + f''_\perp )}{4}
.
\label{localTOK_AB}
\end{align}

Such analysis shows that the high-frequency TO modes at the K point
in the AB structure remain degenerate with a resulting frequency
\begin{eqnarray}
    M \omega^2_{{TO},AB}(K)
    &= &
3 f_\parallel +
\frac{1}{2}f'_\perp + \frac{9}{4} ( f''_\parallel + f''_\perp )
 \hspace{5mm} \mbox{(double degenerate)},
\end{eqnarray}
whereas the interlayer coupled leads to a lift
of the degeneracy in the AA stacking,
resulting in the frequencies
\begin{eqnarray}
    M \omega^2_{{TO},AA}(K)
    &= &
    \left\{
\begin{array}{c}
3 f_\parallel + 3 f''_\parallel 
,
\\
3 f_\parallel +2f'_\perp
3f''_\perp 
.
\end{array}
\right.
 \end{eqnarray}

\subsubsection{High-energy LO/TO modes at the $\Gamma$ point}

Finally, an effective ${\bf k}\cdot {\bf p}$ model can be built
also for the high-energy LO/TO modes at the $\Gamma$ point.
Since the single-layer shows two degenerate modes
at the $\Gamma$ point, also in this case the effective
model will be described by a $4 \times 4$
dynamical matrix resulting by the interlayer coupling
of $2 \times 2$ blocks in each layer.
We can thus write:

\begin{align}
    {{\cal H}}_{LO/TO,\Gamma}^{\alpha}(\vec{k}) &= \left(
    \begin{array}{cc}
     {{\cal H}}_{LO/TO,\Gamma}(\vec{k})+ {\cal V}_{LO/TO,\Gamma}^{\alpha} &
     {{\cal H}}_{LO/TO,\Gamma}^{\alpha} \\
    {{\cal H}}_{LO/TO,\Gamma}^{\alpha,\dagger} & 
    {{\cal H}}_{LO/TO,\Gamma}(\vec{k})
    + {\cal V}_{LO/TO,\Gamma}^{\alpha}
\end{array}
\right),
\end{align}
where ${\cal H}_{LO/LT,\Gamma}(\vec{k})$ is the 
dynamical matrix at the quadratic order
for the single-layer:
\begin{align}
     {{\cal H}}_{LO/TO,\Gamma}(\vec{k})&= \left(
    \begin{array}{cc}
     3 ( f_\parallel + f_\perp ) - \frac{f_\perp ( 3 f_\parallel + f_\perp )}{8 ( f_\parallel + f_\perp )} | \vec{k}|^2 & 0
     \\
     0 & 3 ( f_\parallel + f_\perp ) - \frac{ f_\parallel ( f_\parallel + 3 f_\perp )}{8 ( f_\parallel + f_\perp )} | \vec{k}|^2
\end{array}
\right),
\label{disp_gamma}
\end{align}
${{\cal H}}_{LO/TO,\Gamma}^{\alpha}$, as usual,
is the 
$2 \times 2$ interlayer coupling for each stacking order:
\begin{align}
     {{\cal H}}_{LO/TO,\Gamma}^{AA}
     &= \left(
    \begin{array}{cc}
 - f'_\perp + \frac{\displaystyle 3 ( f''_\parallel + f''_\perp )}{\displaystyle 2} & 0 \\
 0 &  - f'_\perp + \frac{\displaystyle 3 ( f''_\parallel + f''_\perp )}{\displaystyle 2}
\end{array}
\right),
\label{interTOG_AA}
\\
{{\cal H}}_{LO/TO,\Gamma}^{AB}
     &= \left(
    \begin{array}{cc}
 \frac{\displaystyle f'_\perp}{\displaystyle 2} - \frac{\displaystyle 3 ( f''_\parallel + f''_\perp )}{\displaystyle 4} & 0 \\
0 &  \frac{\displaystyle f'_\perp}{\displaystyle 2} - \frac{\displaystyle 3 ( f''_\parallel + f''_\perp )}{\displaystyle 4}
\end{array}
\right),
\label{interTOG_AB}
\end{align}
and ${\cal V}_{LO/TO,\Gamma}^{\alpha}$ describe
the onsite intra-atomic potential:
\begin{align}
     {\cal V}_{LO/TO,\Gamma}^{AA}
     &= \left(
    \begin{array}{cc}
f'_\perp + \frac{\displaystyle 3 ( f''_\parallel + f''_\perp )}{\displaystyle 2} & 0 \\
0 & f'_\perp + \frac{\displaystyle 3 ( f''_\parallel + f''_\perp )}{\displaystyle 2}
\end{array}
\right),
\label{VG_AA}
\\
     {\cal V}_{LO/TO,\Gamma}^{AB}
     &= \left(
    \begin{array}{cc}
\frac{\displaystyle f'_\perp}{\displaystyle 2} + \frac{\displaystyle 9 ( f''_\parallel + f''_\perp )}{\displaystyle 4} & 0 \\
0 & \frac{\displaystyle f'_\perp}{\displaystyle 2} + \frac{\displaystyle 9 ( f''_\parallel + f''_\perp )}{\displaystyle 4}
\end{array}
\right).
\label{VG_AB}
\end{align}

In both stackings, the LO/TO modes at $\Gamma$ in bilayer systems
present two couples of degenerate modes, with frequencies:
\begin{eqnarray}
    M \omega^2_{{LO/TO},AA}(\Gamma)
    &= &
    \left\{ \begin{array}{cc} 
    3 ( f_\parallel + f_\perp ) +3( f''_\parallel + f''_\perp ) & \hspace{5mm} \mbox{(double degenerate)},
    \\
    3 ( f_\parallel + f_\perp )  +2 f'_\perp & \hspace{5mm} \mbox{(double degenerate)},
        \end{array}
    \right.
    \\
    M \omega^2_{{LO/TO},AB}(\Gamma)
    &= &
    \left\{ \begin{array}{cc} 
    3 ( f_\parallel + f_\perp ) + f'_\perp + \frac{3}{2} (f''_\parallel + f''_\perp) & \hspace{5mm} \mbox{(double degenerate)},
    \\
    3 ( f_\parallel + f_\perp ) +3(f''_\parallel + f''_\perp) & \hspace{5mm} \mbox{(double degenerate)}.
    \end{array}
 \right.
\end{eqnarray}

\section{Dynamical matrix in twisted bilayers}

In the above section, we have summarized the effective
${\bf k}\cdot {\bf p}$ model identifying,
for each phonon mode under investigation,
the inter-layer elastic force terms
and the onsite intra-atomic potentials.
A compact view of such local potential is summarized
in Table \ref{table-local} for the relevant degenerate
modes LO/LA at the K point,
LO/TO modes at the $\Gamma$ point,
and TO at the K point.

\begin{table}[t]
\begin{tabular}{|c|c|c|c|c|}
\hline \hline
\multicolumn{5}{c}{LO/LA modes at K} \\
\hline
 stacking & $V_{A,1}$ & $V_{B,1}$ & $V_{A,2}$ & $V_{B,2}$
\\ \hline
AA &
$f'_\perp + 3 ( f''_\parallel + f''_\perp )/2$ &
$f'_\perp + 3 ( f''_\parallel + f''_\perp )/2$ &
$f'_\perp + 3 ( f''_\parallel + f''_\perp )/2$ &
$f'_\perp + 3 ( f''_\parallel + f''_\perp )/2$ 
\\ \hline
AB &
 $f'_\perp + 3 ( f''_\parallel + f''_\perp )/2$ &
$3 ( f''_\parallel + f''_\perp )$ &
$3 ( f''_\parallel + f''_\perp )$ &
$f'_\perp + 3 ( f''_\parallel + f''_\perp )/2$ 
\\ \hline
BA & 
$3 ( f''_\parallel + f''_\perp )$ &
 $f'_\perp + 3 ( f''_\parallel + f''_\perp )/2$ &
 $f'_\perp + 3 ( f''_\parallel + f''_\perp )/2$ &
$3 ( f''_\parallel + f''_\perp )$ 
\\
 \hline \hline
\end{tabular}
\vspace{4mm}
\\
\begin{tabular}{|c|c|c|c|c|}
\hline \hline
\multicolumn{5}{c}{LO/TO modes at $\Gamma$} \\
\hline
 stacking & $V_{x,1}$ & $V_{y,1}$ & $V_{x,2}$ & $V_{y,2}$
\\ \hline
AA &
$f'_\perp + 3 ( f''_\parallel + f''_\perp )/2$ &
$f'_\perp + 3 ( f''_\parallel + f''_\perp )/2$ &
$f'_\perp + 3 ( f''_\parallel + f''_\perp )/2$ &
$f'_\perp + 3 ( f''_\parallel + f''_\perp )/2$ 
\\ \hline
AB &
 $f'_\perp/2 + 9 ( f''_\parallel + f''_\perp )/4$ &
 $f'_\perp/2 + 9 ( f''_\parallel + f''_\perp )/4$ &
$f'_\perp/2 + 9 ( f''_\parallel + f''_\perp )/4$ &
$f'_\perp/2 + 9 ( f''_\parallel + f''_\perp )/4$ 
\\ \hline
BA & 
 $f'_\perp/2 + 9 ( f''_\parallel + f''_\perp )/4$ &
 $f'_\perp/2 + 9 ( f''_\parallel + f''_\perp )/4$ &
 $f'_\perp/2 + 9 ( f''_\parallel + f''_\perp )/4$ &
 $f'_\perp/2 + 9 ( f''_\parallel + f''_\perp )/4$ 
\\
 \hline \hline
\end{tabular}
\vspace{4mm}
\\
\begin{tabular}{|c|c|c|}
\hline \hline
\multicolumn{3}{c}{TO modes at K} \\
\hline
 stacking & $V_{1}$ & $V_{2}$ 
\\ \hline
AA & 
$f'_\perp + 3 ( f''_\parallel + f''_\perp )/2$ &
$f'_\perp + 3 ( f''_\parallel + f''_\perp )/2$ 
\\ \hline
AB &
 $f'_\perp/2 + 9 ( f''_\parallel + f''_\perp )/4$ &
 $f'_\perp/2 + 9 ( f''_\parallel + f''_\perp )/4$ 
\\ \hline
BA & 
 $f'_\perp/2 + 9 ( f''_\parallel + f''_\perp )/4$ &
 $f'_\perp/2 + 9 ( f''_\parallel + f''_\perp )/4$ 
\\
 \hline \hline
\end{tabular}
\caption{Onsite atomic potentials in different stacking orders
of bilayer systems for modes LO/LA at the K point
LO/TO at the $\Gamma$ point, and TO at the K point.
For the LO/LA modes, the potential is specified for
each sublattice of each layer; for the
LO/TO modes is specified for $x-$, $y$-eigenvectors
in Eq. (\ref{eigenTOLO}) 
of each layer;
for the TO mode at K is specified for the
TO eigenvector in Eq. (\ref{etok}) in each layer.
}
\label{table-local}
\end{table}

Equipped with the knowledge of the role of the interlayer coupling
for the characteristic bilayer structures AA and AB,
we can now estimate the dynamical matrix for twisted bilayer systems within
the framework of a continuum model.

The analysis follows slightly different procedures
for each mode, accounting for the different
characteristic vectors (K vs. $\Gamma$),
and for the different size of the Hilbert space
[doublet modes for LO/LA(K) and LO/TO($\Gamma$)
vs. single non degenerate mode for TO(K)]

\subsubsection{Dirac LO/LA modes at the K point}

The continuum model for the phonon Dirac spinor
associated with the LO/LA modes at the K points
follows a standard approach as employed for
the electronic dispersion.
Within this framework,
we consider first two decoupled single-layer systems
in the AA stacking, upon which we apply a twist with angle $\theta$.
Such geometric configuration defines three characteristic momenta
${\bf Q}_1=k_\theta(0,1)$, ${\bf Q}_2=k_\theta(\sqrt{3}/2,1/2)$, 
${\bf Q}_3=k_\theta(-\sqrt{3}/2,1/2)$, where
$k_\theta=2k_{\rm BZ}\sin(\theta/2)$,
$k_{\rm BZ}$ being the absolute value of the momentum of the Brillouin zone edge.
The ${\bf Q}_\nu$ momenta rule the relevant
tunneling processes between layers $\alpha$ and $\beta$ by means
of the interlayer couplings:
\begin{eqnarray}
T^{\alpha\beta}({\bf r})
&=&
\bar{t}
\sum_\nu
T_\nu^{\alpha\beta}
\mbox{e}^{i {\bf Q}_\nu \cdot {\bf r}}
,
\end{eqnarray}
where as usual (assuming translational invariance with
respect to the relative shift of the two layers)
\begin{eqnarray}
\hat{T}_1
&=&
\left(
\begin{array}{cc}
1 & 1 \\
1 & 1
\end {array}
\right)
,
\hspace{5mm}
\hat{T}_2
=
\left(
\begin{array}{cc}
\mbox{e}^{-2\pi i /3} & 1 \\
\mbox{e}^{2\pi i /3} & \mbox{e}^{-2\pi i /3}
\end {array}
\right)
,
\hspace{5mm}
\hat{T}_3
=
\left(
\begin{array}{cc}
\mbox{e}^{2\pi i /3} & 1 \\
\mbox{e}^{-2\pi i /3} & \mbox{e}^{2\pi i /3}
\end {array}
\right)
.
\label{tmatrices}
\end{eqnarray}
Following the procedure in Ref. \cite{LPN07},
the parameter 
\begin{eqnarray}
\bar{t}
&=&
\sqrt{
t_{AA,i}^2+t_{AB,i}^2}
,
\end{eqnarray}
is here an effective energy scale
obtained
by interpolating the AA and the AB/BA 
interlayer coupling matrices
${{\cal H}}_{LO/LA}^{AA}$ and
${{\cal H}}_{LO/LA}^{AB}$.

More in particular,
following a standard procedure based
on a perturbation analysis,
using Eqs. (\ref{interTOTA_AA})-(\ref{interTOTA_AB}).
we get:
\begin{eqnarray}
t_{AA,LO/LA}(K)
&=&
-\frac{f'_\perp}{3}
,
\label{LOLA-perpAA}
\\
t_{AB,LO/LA}(K)
&=&
-\frac{1}{2}
(f''_\parallel - f''_\perp)
.
\label{LOLA-perpAB}
\end{eqnarray}



Besides the interlayer tunnelling processes,
the interlayer elastic coupling between the
twisted layers gives rise
for each mode
to local atomic potentials
which are described by the diagonal matrices
${{\cal V}}_{\alpha}^{AA}$ and
${{\cal V}}_{\alpha}^{AB}$,
${{\cal V}}_{\alpha}^{BA}$,
as they are summarized in Table \ref{table-local}
for the AA and AB structures.
In twisted bilayer systems,
these potentials can be expanded in reciprocal lattice vectors,
in the same way as electrostatic potentials are incorporated into the continuum model of the
electron band structure of twisted bilayer graphene\cite{GW18}.

More in particular, for each mode $\alpha$
we can define an average potential $\bar{V}^{AB}_\alpha$
and a potential difference $\Delta V^{BA}_\alpha$
for the AB and BA structures,
\begin{eqnarray}
\bar{V}^{AB}_\alpha
&=&
\frac{V^{AB}_\alpha+V^{BA}_\alpha}{2}
,
\\
\Delta V^{AB}_\alpha
&=&
\frac{V^{AB}_\alpha-V^{BA}_\alpha}{2}
.
\label{inplanep}
\end{eqnarray}
Furthermore we can define a potential difference between
the average potential in the AA and AB regions:
\begin{eqnarray}
\Delta V_\alpha
&=
V^{AA}_\alpha
-
\bar{V}^{AB}_\alpha
.
\end{eqnarray}
Such difference between $AA$ and $AB$ regions can be described in terms of moir\'e harmonics.
Expanding into the first star of moir\'e reciprocal lattice vectors, we obtain:
\begin{align}
V_\alpha ( \vec{r} ) &= \frac{\Delta V_\alpha}{9} \sum_{i =1,2,3} \cos ( \vec{G}_i \vec{r} )
.
\label{pot1}
\end{align}
In similar way,
the layer dependent modulation of the potentials at the $AB$ and $BA$ regions can be written as:
\begin{align}
    \Delta V_\alpha ( \vec{r} ) &= \pm \frac{2 \Delta V^{AB}_\alpha}{3 \sqrt{3}} \sum_{i =1,2,3} \sin ( \vec{G}_i \vec{r} )
    \label{pot2}
\end{align}
The potentials in Eqs. (\ref{pot1})-(\ref{pot2}) can be incorporated into a continuum model in the same way as the electrostatic potential are added to the electronic continuum model of twisted bilayer graphene.\cite{GW18}
We keep only the first star of reciprocal lattice vectors, and include these potentials into the continuum model in a similar way to the inclusion of the (scalar but sublattice independent) Hartree potential in Ref. \cite{GW18}.

\begin{figure*}[t]
    \includegraphics[width=1.0\textwidth]{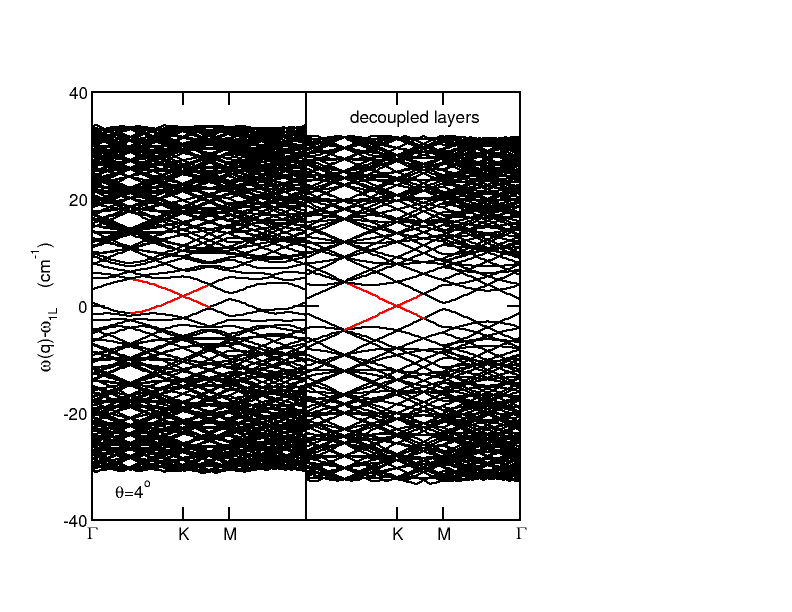}
    \caption{Band dispersion for the LO/LA modes in the reduced moir\'e Brillouin zone for twisted bilayer case at $\theta=4^\circ$(left panel)
    and for the reference case of the decoupled layers (right panel).
    Marked in red is the linear Dirac-like dispersion.
    }
    \label{f-comparison}
\end{figure*}

\begin{figure*}[t]
    \includegraphics[width=0.6\textwidth]{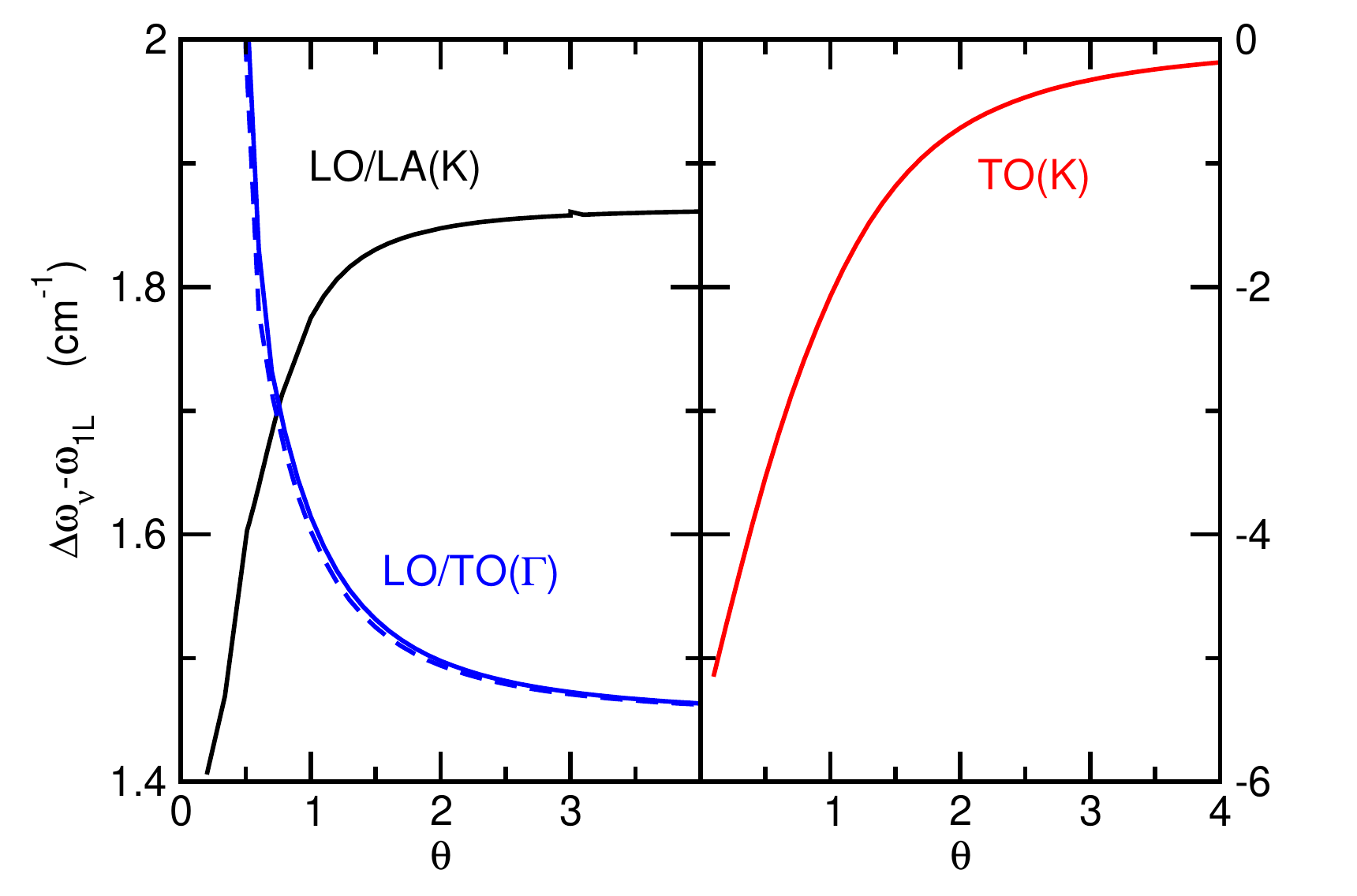}
    \caption{Angle dependence of the band shift $\Delta\omega_\nu$
    for the relevant modes $\nu=LO/LA(K), TO(K), LO/TO(\Gamma)$.
    For the LO/TO modes, the blue solid line refers to the even symmetry,
    whic the dashed line to the odd symemtry.
    }
    \label{f-shifts}
\end{figure*}

The full phonon dispersion of the LO/LA modes
in the reduced moir\'e Brillouin zone
can be thus computed, as shown for instance
in Fig. 2a,b of the main text.
A comparison between the LO/LA dispersion in the twisted
case with $\theta=4^\circ$ and the reference case of two decoupled layers
is shown in Fig. \ref{f-comparison}.
The Dirac bands (marked in red color) are easily identified.
One can notice two main features: an overall upwards energy shift
$\Delta\omega_{LO/LA}$
of the main dispersion, of about $\sim 1.8$ cm$^{-1}$;
and a renormalization of the linear Dirac dispersion,
just like for the electronic case.
For given angle, we evaluate the coefficient $v^*$ of the linear dispersion
$\hbar\omega_{LO/LA}(\vec{k})
=
\hbar\omega_{LO/LA}(K)\pm v^* |\vec{p}|$
of the Dirac mode in the twisted case in comparison
with the linear coefficient $v$ of the uncoupled single-layer.
The parameter $R=v^*/v$ provides thus the ``renormalization'' band
factor of the twisted Dirac phonons dispersion,
in similar was as shown in the inset of Fig. 4 in Ref. \cite{BM11}.
The angle dependence of the renormalization band factor
is shown in Fig. 2c of the main text, showing a remarkable
trend toward flat Dirac bands for $\theta \lesssim 2^\circ$.
Also interesting, is the analysis
of the phonon band-shift $\Delta\omega_{LO/LA}$,
reported in Fig. \ref{f-shifts}, which shows
a negligible dependence on $\theta$ for large twiste angles,
but a sizable drop at low angle, in the region where the band renormalization
is also more marked.
Such angle dependence of the phonon band-shift $\Delta\omega_{LO/LA}$,
is expected to be reflected in an observable angle-dependence mode softening.

If we neglect for the moment the role of
the onsite potentials,
we can use a perturbative analysis of the interlayer tunneling terms (see for instance Ref. \onlinecite{BM11})
in order to obtain an estimate of the first ``magic''
angle at which the phonon dispersion
of the twisted bilayer vanishes.

For the case of the Dirac-like phonons LO/LA at
the K point, such condition occurs when
\begin{align}
    2 \sin\left( \frac{\bar{\theta}_{LO/LA}}{2} \right)  &= \frac{3}{4 \pi} \frac{\sqrt{3 \bar{t}^2}}{V} =
     \frac{\sqrt{3}}{\pi( f_\parallel - f_\perp )} 
     \sqrt{3 \left[ (f')_\perp^2 + \frac{9 \left( f''_\parallel - f''_\perp \right)^2}{4} \right]}
     .
\end{align}
Using the force-constant parameters
extracted from the comparison with ab-initio
calculations (see next Section), 
this equation gives $\bar{\theta}_{LO/LA}(K) \approx 2.1^\circ$.

\subsubsection{High-energy TO modes at the K point}

The TO phonons at the K point in twisted systems
can also be described as a variation of the continuum model
discussed in Ref. \onlinecite{LPN07},
where we expand the dynamical matrix in Bloch wavefunctions at each layer. 
Just as in the previous case, the interlayer scattering is
still governed by the
three characteristic momenta
${\bf Q}_1=k_\theta(0,1)$, ${\bf Q}_2=k_\theta(\sqrt{3}/2,1/2)$, 
${\bf Q}_3=k_\theta(-\sqrt{3}/2,1/2)$.
With respect to the LO/LA modes,
there are however two main differences.
One one hand, the phonon dispersion in the singe-layer
does not obey to a linear behavior but to a 
quadratic one as shown in Eq.(\ref{disp_TO}).
On the other hand,
as this mode in single-layer is not degenerate,
the $2 \times 2$ spinor structure of Eq. (\ref{tmatrices}) is lifted,
and for each momentum only one wave function per layer is needed.
The terms which define the interlayer tunneling are thus
reduced to $1 \times 1$ numbers, as in
Eq. (\ref{interTOK_AA})-(\ref{interTOK_AB}).
A block state in one layer is coupled by these terms to three Bloch states in the second layers, and the three terms just acquire a phase modulation, $e^{i \phi_j} , j = 1,2,3$ (see Ref. \onlinecite{WLTMM19} for a related approach to the electronic states in twisted homo-bilayer dichalcogenides).
Finally we include the intra-layer sublattice potentials $V_1 , V_2$
as shown in Table \ref{table-local}. Just as in the case of the Dirac LO/LA modes, these local potentials are expanded using the first star of reciprocal lattice vectors. 

The phonon dispersion so computed is shown in panels d-e of Fig. 2 of the main text.
Also in this case, a renormalization band factor can be
evaluate by the ratio of the quadratic dispersion
$\hbar\omega_{TO}(\vec{k})
=
\hbar\omega_{TO}(K)+ \alpha^* |\vec{p}|^2$
in the twisted bilayer and in the single layer cases, $R=\alpha^*/\alpha$.
The angle dependence of the such renormalization band
factor is also shown in panel f of Fig. 2 of the main text,
and the angle dependence of
the band-shift $\Delta\omega_{TO}$ for this mode in Fig. \ref{f-shifts}.

Also for the TO modes at the K point,
using the same perturbative approach for the interlayer tunneling, 
we can provide a qualitative estimate of the magic
angle were
the positive quadratic dispersion at K is renormalized to zero.
Using the appropriate expression for a quadratic dispersion we find:
\begin{align}
    2 \sin\left( \frac{\bar{\theta}_{TO}}{2} \right)  &= \frac{3}{4\pi}
    \times \sqrt{\frac{2 \sqrt{3} \left| f'_\perp + \frac{3 \left( f''_\perp - f''_\parallel \right)}{2} \right| \left( f_\parallel - f_\perp \right)}{f_\parallel f_\perp} }
    .
    \end{align}
Note that this result depends only
on the prefactor of the quadratic dispersion of the TO mode of a monolayer, and on the set of interlayer force-constant parameters
which describes the force between layers in the AA structure,
as defined in Eq.(\ref{interTOK_AA}).

Using the force-constant parameters
extracted from the comparison with ab-initio
calculations, we get
$\bar{\theta}_{TO}(K) \approx 1.0^\circ$.

\subsubsection{High-energy LO/TO modes at the $\Gamma$ point}

The analysis of the twisting on the LO/TO modes at the $\Gamma$
point is somehow much simpler than the previous cases discussed
for phonons at the K point.
The phonons at the $\Gamma$ point in the individual layers of a twisted bilayer are indeed mapped onto the $\Gamma$ point of the twisted bilayer Brillouin zone, unlike the modes at the $K$ point, which
are instead mapped onto the K or K$^\prime$ point of the twisted bilayer $\Gamma$ point, depending on the layer index.
Moreover, as shown in Eqs. (\ref{interTOG_AA}), (\ref{interTOG_AB}),(\ref{VG_AA}), (\ref{VG_AB}), and in Table \ref{table-local}, all the interlayer coupling terms are just multiple of the unit matrix
in the $2 \times 2$ space, as specified in Eqs. (\ref{interTOG_AA})-(\ref{interTOG_AB}).
 Hence, the LO and TO modes can be treated independently. The continuum model for the phonons at the $\Gamma$ point can be written in terms of Bloch waves defined on the reciprocal lattices for each layer, where the two lattices lie on top of each other, unlike the case of the modes at $K$, where the two reciprocal lattices are displaced with respect to each other.\cite{BM11} A similar model has been employed for electrons near the $\Gamma$ point of the valence band in transition metal dichalcogenides.\cite{ZLF21,AM21}
 Furthermore, the interlayer coupling terms allow for the separation of the two LO and the two TO modes into even and odd combinations which also are decoupled, so that the continuum model for the
 dynamical matrix of the phonons at $\Gamma$ can be split into four independent blocks.

Using such procedure,
the computed phonon dispersion and the angle dependence of the renormalization
band factor for the TO modes close to the $\Gamma$ point
are shown in panels g-i of Fig. 2 of the main text.
We also show in Fig. \ref{f-shifts}
the angle dependence of
the band-shifts $\Delta\omega_{LO/TO}$
of these modes for both even and odd symmetries.
Note that, since LO and TO are degenerate at the $\Gamma$ point,
a similar band-shift applies for both LO and TO phonon dispersions.

Obeying to the above simplifications,
also the estimates of magic angles where
the dispersion of the TO/LO modes
at the $\Gamma$ point is renormalized to zero ($R \to 0$)
is significantly simplified.

Since, as discussed above, the continuum model for the phonons at $\Gamma$ can be split into four independent hamiltonian blocks,
corresponding to even and odd layer combinations,
we obtain {\em four} magic angles, one for each
LO vs. TO and even vs. odd combinations.
Using the interlayer coupling reported
in Eqs. (\ref{interTOG_AA})-(\ref{interTOG_AB}).
the quadratic dispersion of Eq.(\ref{disp_gamma}),
and using non local coupling between Bloch waves separated by a reciprocal lattice vector in the first star, 
the perturbation theory results in the following relations:
\begin{align}
2 \sin\left( \frac{\bar{\theta}}{2} \right) &=   \frac{\sqrt{3}}{4 \pi} \sqrt{\frac{| \tilde{f}_1 \pm \tilde{f}'_1 |}{v_2^{LO}}} 
,
\nonumber \\
2 \sin\left( \frac{\bar{\theta}}{2} \right) &=   \frac{\sqrt{3}}{4 \pi} \sqrt{\frac{| \tilde{f}_1 \pm \tilde{f}'_1 |}{v_2^{TO}}} 
,
\end{align}
where:
\begin{align}
\tilde{f}_1 &=  \frac{f'_\perp - \frac{3}{2} \left( f''_\parallel + f''_\perp \right)}{18}
,
\nonumber \\
\tilde{f}'_1 &=  \frac{- 3 f'_\perp+ \frac{3}{2} \left( f''_\parallel + f''_\perp \right)}{18}
,
\nonumber \\
  v_2^{LO} &=  \frac{f_\perp ( 3 f_\parallel + f_\perp )}{8 ( f_\parallel + f_\perp )}, \nonumber \\
    v_2^{TO} &= \frac{ f_\parallel ( f_\parallel + 3 f_\perp )}{8 ( f_\parallel + f_\perp )} .
\end{align}

Using the force-constant parameters
extracted from the comparison with ab-initio
calculations, we find
$\bar{\theta}_{LO}(\Gamma,+) \approx 0.44^\circ$,
$\bar{\theta}_{TO}(\Gamma,+) \approx 0.42^\circ$,
$\bar{\theta}_{LO}(\Gamma,-) \approx 0.83^\circ$,
$\bar{\theta}_{TO}(\Gamma,-) \approx 0.79^\circ$.

\section{Mapping ab-initio calculations onto force-constant model}

In order to achieve a realistic modelling of the
lattice dynamics in twisted bilayer graphene,
we use ab-initio calculations in order
to extract the appropriate parameters for the
force-costant model.

Density functional theory calculations (DFT) were performed using Quantum Espresso (QE)\cite{Giannozzi2009,Giannozzi2017,Giannozzi2020}.
For the electronic calculations, we use the Generalized Gradient Approximation (GGA), especifically, the functional of Perdew, Burke and Ernzerhof \cite{PBE96}.
We set the energy cutoff for the wavefunctions to 240 Ry and the cutoff for the density to 1400 Ry.
In order to obtain the correct value for the interlayer spacing in the case of bilayer graphene, we use the Grimme approximation\cite{grimme2006semiempirical}.
The Brillouin zone was sampled using the Monkhorst–Pack
scheme \cite{MonPac76} with a grid of $32\times32\times1$ k-points.
We have optimized the lattice vectors and relaxed the atomic positions to forces lowers than 1 eV/\AA.
The phonon band structure was calculated using Density Functional Perturbation Theory (DFPT)\cite{baroni2001} as implemented in QE.

The force-constant (FC) model here employed for the phonon dispersion
in bilayer graphene depends on {\em five} independent
elastic parameters, i.e.
$f_\parallel$, $f_\perp$,
$f_\perp^\prime$,
$f_\parallel^{\prime\prime}$, and $f_\perp^{\prime\prime}$.
Given the pivotal role in our discussion of the
Dirac-like LO/LA modes at the K point of single-layer
and bilayer structures, we calibrate our FC parameters
in order to reproduce in the best way
these Dirac-like features.

A first crucial feature is in the single-layer
the Dirac-like linear dispersion of the LO/LA
modes at the K point,
\begin{eqnarray}
\hbar\omega_{LO/LA}(\vec{k})
&=&
\hbar\omega_{LO/LA}(K)\pm v |\vec{p}|
,
\end{eqnarray}
where $\vec{p}=\vec{k}-K$.
Our DFT calculations 
find
\begin{eqnarray}
v
&=&
7.25 \times 10^{4}\, \mbox{cm/s} = 4.77\,\, \mbox{meV \AA}
.
\end{eqnarray}
Further relevant ab-initio inputs
are the LO/LA frequencies in the single-layer
as well as AA and AB stackings.
Their values are reported in Table \ref{tab:dft2}
\begin{table}[t]
    \centering
    \begin{tabular}{c|c|c}
    \hline
    \hline
1L & AA & AB \\
    \hline
    1215.34$^*$ & 1214.78$^*$ & 1215.41 \\
            & 1217.41$^*$ & 1215.54$^*$ \\
            &             & 1216.26
\\
    \hline
    \hline
    \end{tabular}
    \caption{Ab-initio LO/LA phonon frequencies at the K point
    in units of cm$^{-1}$. Frequencies marked with ($*$)
    are double degenerate.}
    \label{tab:dft2}
\end{table}

These first-principle inputs can be now employed
in order to estimate proper force-constant parameters.
More in details,
from the relation:
\begin{eqnarray}
M\omega^2_{LO/LA}(K)
&=&
\frac{3}{2}(f_\parallel+f_\perp),
\end{eqnarray}
we get the value of the linear combination $f_\parallel+f_\perp$:
\begin{eqnarray}
f_\parallel+f_\perp
&=&
43.85 \mbox{\,\,eV/\AA}^2.
\end{eqnarray}
The first-principles value of the Dirac velocity $v$
of these modes close to K provide further
analytical constraints.
From Eq. (\ref{dynDirac}), which refers to the
{\em dynamical matrix},
we can obtain an analytical expression for $v$:
\begin{eqnarray}
v
&=&
\sqrt{\frac{3(f_\parallel+f_\perp)}{2}}
\frac{f_\parallel-f_\perp}{f_\parallel+f_\perp}
\frac{\hbar a}{4\sqrt{3}}
,
\end {eqnarray}

Using these inputs we can determine
thus the values of
the in-plane force-constant parameters
$f_\parallel$ and $f_\perp$.

The interlayer force-constant parameters
$f_\perp^\prime$,
$f_\parallel^{\prime\prime}$, and $f_\perp^{\prime\prime}$
can be estimated from the spectrum
of the LO/LA modes at the K point in the AA and AB
structures.
Using Eqs. (\ref{LOLAsplitAA}),   
from the splitting of the computed frequencies
of the Dirac LO/LA modes in the AA structure,
we can extract the value of $2f_\perp^\prime$.
In similar way, using Eqs. (\ref{LOLAsplitAB}),
we can extract the linear combination
 $f_\parallel^{\prime\prime}-f_\perp^{\prime\prime}$
from the splitting of the single-degenerate
LO/LA levels at K in the AB structure.
In order to have a complete set of force-constant parameters,
we have to further extract from the ab-initio calculations
the linear combination 
$f_\parallel^{\prime\prime}+f_\perp^{\prime\prime}$.
This can be obtained, using Eqs. (\ref{LOLA_1L}) and (\ref{LOLAsplitAA}),
by comparing the frequency shift of the symmetric LO/LA modes in the AA bilayer
with respect to the reference frequency of the LO/LA modes
in the single-layer case.
The force-constant parameters
so extracted from the ab-initio input are listed
in Table \ref{tab-param}.

\begin{table}[b]
    \centering
    \begin{tabular}{c|c|c|c|c}
    \hline
    \hline
$f_\parallel$ & $f_\perp$ &
$f_\perp^\prime$ &
$f_\parallel^{\prime\prime}$ & $f_\perp^{\prime\prime}$
 \\
$23.882$ & $19.973$ & $-0.143$ & $0.090$ & $0.059$ 
\\
    \hline
    \hline
    \end{tabular}
    \caption{Force-constant parameters extracted
    from ab-initio calculations, in units of eV/\AA$^2$.}
    \label{tab-param}
\end{table}

\begin{figure*}[t]
    \centering
    \includegraphics{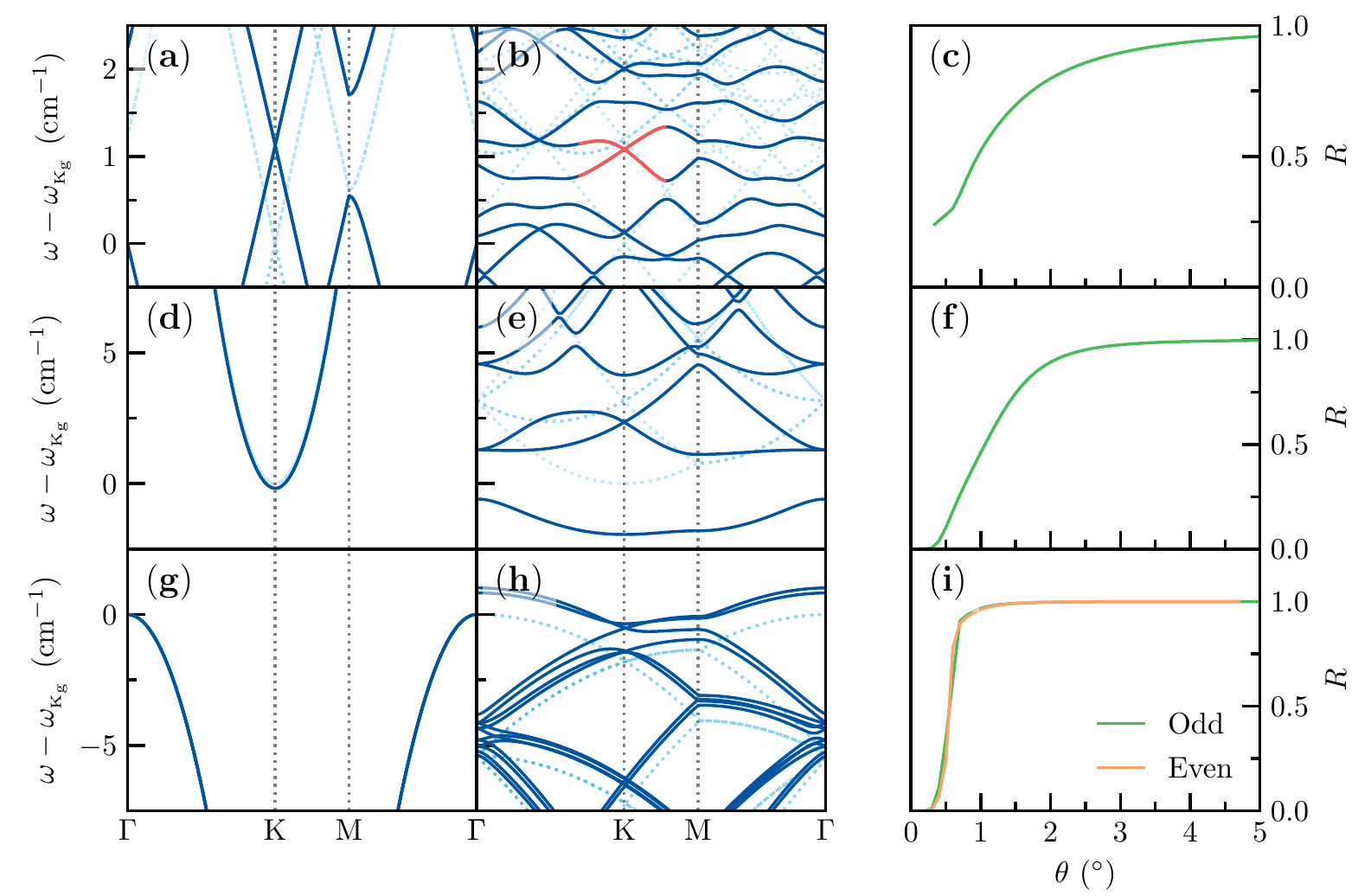}
    \caption{(
    Phonon dispersion in the moir\'e Brillouin zone
    for $\theta=4.0^\circ$ and $\theta=1.05^\circ$ (left and middle columns)
and phonon band renormalization factor $R$ as function of the twist angle $\theta$
computed by using the  force-parameter set
    $f_\parallel=23.882$ eV/\AA$^2$, $f_\perp=19.973$ eV/\AA$^2$,
$f_\perp^\prime=-0.143$ eV/\AA$^2$,
$f_\parallel^{\prime\prime}=0.070$ eV/\AA$^2$,
$f_\perp^{\prime\prime}=0.039$ eV/\AA$^2$.
    }
    \label{fig:bands_twist_alt}
\end{figure*}

It is worth to mentioning that,
while the parameters 
$f_\parallel$, $f_\perp$,
$f_\perp^\prime$,
and the linear combination 
$f_\parallel^{\prime\prime}-f_\perp^{\prime\prime}$
are extracted in a compelling way from the analysis
of the LO/LA modes at the K point in single-layer and bilayer structure,
the determination of the last condition,
namely 
$f_\parallel^{\prime\prime}+f_\perp^{\prime\prime}$,
is less univocally.
As an alternative procedure, we could estimate
the linear combination $f_\parallel^{\prime\prime}+f_\perp^{\prime\prime}$,
using Eq. (\ref{LOLAsplitAB}),
from the analysis of the  
the relative frequency shift of the LO/LA modes {\em in the AB bilayer}
with respect to the reference frequency of the LO/LA modes
in the single-layer case.
Along this derivation, one would extract slightly different
values of $f_\parallel^{\prime\prime}$, $f_\perp^{\prime\prime}$,
namely
$f_\parallel^{\prime\prime}=0.070$ eV/\AA$^2$,
$f_\perp^{\prime\prime}=0.039$ eV/\AA$^2$.
Note however that such slight uncertainty on the quantity
$f_\parallel^{\prime\prime}-f_\perp^{\prime\prime}$
does not affect sensibly the twisted phonon dispersion
since the relevant interlayer tunneling processes,
for the LO/LA modes at the K point,
as well as for the TO at K and TO/LO at $\Gamma$,
are essentially ruled 
[see Eqs. (\ref{LOLA-perpAA}), (\ref{LOLA-perpAB})
(\ref{interTOK_AA}), (\ref{interTOK_AB}),
(\ref{interTOG_AA}), and (\ref{interTOG_AB})]
only by the parameters $f_\perp^\prime$,
and $(f_\parallel^{\prime\prime}-f_\perp^{\prime\prime})$
which can be determined without ambiguity
from the first-principle calculations.
In Fig. \ref{fig:bands_twist_alt} we show
the phonon dispersions for $\theta=1.05^\circ, 4^\circ$
and the angle dependence of the band renormalization
factor $R$ for
the force-constant parameter set with
$f_\parallel^{\prime\prime}=0.070$ eV/\AA$^2$,
$f_\perp^{\prime\prime}=0.039$ eV/\AA$^2$.
Both features appear essentially identical to the
results shown in the main text with the parameters
listed in Table \ref{tab-param}.

\clearpage
\bibliography{biblioart}
 


\end{document}